\def\la{\mathrel{\mathchoice {\vcenter{\offinterlineskip\halign{\hfil
$\displaystyle##$\hfil\cr<\cr\sim\cr}}}
{\vcenter{\offinterlineskip\halign{\hfil$\textstyle##$\hfil\cr
<\cr\sim\cr}}}
{\vcenter{\offinterlineskip\halign{\hfil$\scriptstyle##$\hfil\cr
<\cr\sim\cr}}}
{\vcenter{\offinterlineskip\halign{\hfil$\scriptscriptstyle##$\hfil\cr
<\cr\sim\cr}}}}}

\def\ga{\mathrel{\mathchoice {\vcenter{\offinterlineskip\halign{\hfil
$\displaystyle##$\hfil\cr>\cr\sim\cr}}}
{\vcenter{\offinterlineskip\halign{\hfil$\textstyle##$\hfil\cr
>\cr\sim\cr}}}
{\vcenter{\offinterlineskip\halign{\hfil$\scriptstyle##$\hfil\cr
>\cr\sim\cr}}}
{\vcenter{\offinterlineskip\halign{\hfil$\scriptscriptstyle##$\hfil\cr
>\cr\sim\cr}}}}}

\def\p {$\pm$}

\def\solmass {$\hbox{M}_\odot$}

\def\kms {\hbox{${\rm km\, s}^{-1}$}} 
\def\cmsq  {$\hbox{{\rm cm}}^{-2}$}    
\def\arcsec {\hbox{$^{\prime\prime}$}}

\def\percc {$\hbox{{\rm cm}}^{-3}$}    
\def\MOLH {\hbox{${\rm H}_2$}}  
\def\MOLN {\hbox{${\rm N}_2$}}  
\def\AMM {\hbox{${\rm NH}_{3}$}} 
\def\HCOP {\hbox{${\rm HCO}^+$}}      
\def\HTHCOP {\hbox{${\rm H^{13}CO}^+$}}      
\def\HCEIOP {\hbox{${\rm HC^{18}O}^+$}}      
\def\HCSEOP {\hbox{${\rm HC^{17}O}^+$}}      
\def\DCOP {\hbox{${\rm DCO}^+$}}    
\def\DTHCOP {\hbox{${\rm D^{13}CO}^+$}}    
\def\HTHP {\hbox{${\rm H}_{3}^{+}$}}   
\def\HTDP {\hbox{${\rm H}_{2}{\rm D}^{+}$}}   
\def\HTHOP {\hbox{${\rm H}_{3}{\rm O}^{+}$}} 
\def\TWCO {\hbox{$^{12}{\rm CO}$}}   
\def\THCO {\hbox{$^{13}{\rm CO}$}}   
\def\CEIO {\hbox{${\rm C}^{18}{\rm O}$}}   
\def\CSEO {\hbox{${\rm C}^{17}{\rm O}$}}   
\def\NTHP {\hbox{${\rm N}_2{\rm H}^+$}} 
\def\NTDP {\hbox{${\rm N}_2{\rm D}^+$}} 
\def\CYAC {\hbox{${\rm HC}_3{\rm N}$}}     
\documentstyle[11pt,aaspp4]{article}

\lefthead{Caselli et al.}
\righthead{The ionization degree in L1544}

\begin{document}

\title{Molecular ions in L1544. II. The ionization degree}

\author{P. Caselli , C. M. Walmsley, \and A.Zucconi}
\affil{Osservatorio Astrofisico di Arcetri, Largo E. Fermi 5, I-50125
Firenze, Italy; caselli@arcetri.astro.it; walmsley@arcetri.astro.it;
zucconi@arcetri.astro.it}

\author{M. Tafalla}
\affil{Observatorio Astron\'omico Nacional (IGN), Campus Universitario,
E-28800 Alcal\'a de Henares (Madrid), Spain; tafalla@oan.es}

\author{L. Dore}
\affil{Dip. Chimica "Ciamincian", Universit\`{a} di Bologna,
Via Selmi 2, I-40126, Bologna, Italy; dore@ciam.unibo.it}

\and

\author{P. C. Myers}
\affil{Harvard--Smithsonian Center for Astrophysics, MS 42, 60
Garden Street, Cambridge, MA 02138, U.S.A.; pmyers@cfa.harvard.edu}




\begin{abstract}

 The maps presented in Paper I are here used to infer the
 variation of the column densities of \HCOP , \DCOP , \NTHP , and
 \NTDP \ as a function of distance from the dust peak.  These results are
 interpreted with the aid of a crude chemical model which predicts the
 abundances of these species as a function of radius in a spherically
 symmetric model with radial density distribution inferred from
 the observations of dust emission at millimeter wavelengths and dust
 absorption in the infrared.  Our main observational finding is that
 the $N(\NTDP )/N(\NTHP )$ column density ratio is of order 0.2 towards the 
 L1544 dust peak as
 compared to $N(\DCOP )/N(\HCOP )$ = 0.04.  We conclude that this
 result as well as the general finding that \NTHP and \NTDP \ correlate
 well with the dust is caused by CO being depleted to a much higher
 degree than molecular nitrogen in the high density core of L1544.
 Depletion also favors deuterium enhancement and thus \NTDP , which 
traces the dense and highly CO--depleted core nucleus, is
 much more enhanced than \DCOP .  Our models do not uniquely define
 the chemistry in the high density depleted nucleus of L1544 but they
 do suggest that the ionization degree is a few times $10^{-9}$ and that
 the ambipolar diffusion time scale is locally similar to the free
 fall time.  It seems likely that the lower limit which one obtains to
 ionization degree by summing all observable molecular ions is not
 a great underestimate of the true ionization degree. We predict that 
atomic oxygen is abundant in the dense core and, if
 so,  \HTHOP \ may be the main ion in the central highly depleted region
of the core.

\end{abstract}


\keywords{ISM: individual (L1544) -- ISM: dust, extinction -- ISM: molecules}


%

\section{Introduction}

The ionization degree $x(e)$ (= $n(e)/n(\rm H_2)$, with $n(e)$ and $n(\rm
H_2)$ the electron and H$_2$ number density, respectively)
in dense molecular clouds plays a key role in
 determining the initial conditions which precede the collapse to
 form a star (e.g. Shu, Adams \& Lizano 1987). 
If the magnetic field threading the dense gas is
 sufficiently large to prevent immediate collapse, ambipolar
 diffusion of neutrals across field lines can lead to a situation
 where a dense core of gas is gravitationally unstable. The time scale
 for such ambipolar diffusion is  proportional to the
 ionization degree and it therefore becomes of interest to
 develop methods of estimating the ionization degree based upon the
 abundances of various species which trace the
 ionization in the dense gas.

 In an earlier article (\cite{CWT98}, hereafter CWT98), we studied the
possibility of using the [\DCOP]/[\HCOP] abundance ratio as a tracer of the
 ionization degree in dense molecular gas, following previous work by
 \cite{WSG79} and \cite{GLW82}.   We used the time--dependent
 models of ~\cite{LBH96} to study the expected behavior of
 a variety of species thought to be relevant in this context and
 compared with the abundances measured by ~\cite{BLL95}.
 An analogous study carried out by ~\cite{WBC98} reached
 very similar results. We confirmed earlier
 work (\cite{DL84}) which showed that a critical parameter
 in addition to the ionization degree which determines the
 fractionation of deuterated species (and hence abundance ratios
 such as [\DCOP]/[\HCOP] ) is the degree of depletion of carbon and
 oxygen bearing species onto the surfaces of interstellar dust grains.
 Measuring the ionization degree in reliable fashion therefore requires
 the use of observations to solve for both $x(e)$ and depletion.
 One possible approach to this problem discussed
 by CWT98 was to use the cyanoacetylene (\CYAC ) abundance
 as a ``depletion indicator'' (see also ~\cite{RHT97}).  A more
detailed analysis on the possibility to use \CYAC \ to estimate
the degree of depletion in a dense core is currently under way
(Comito et al., in prep.).

 The CWT98 and ~\cite{WBC98} studies used  data with
 angular resolution of the order of an arc minute and  did not
 search for spatial variations in the ionization degree and
 hence in the [\DCOP]/[\HCOP] ratio.  Studies of variations
in the deuterium fractionation and in the degree of ionization across 
cores have been previously done in active star forming regions such as
the R Coronae Australis molecular cloud (\cite{ACH99}), and in massive
cores located in regions that are currently forming stellar clusters 
(\cite{BPW99}).  In the former case, 
the electron fraction was found to decrease
near the cluster of newly born stars, whereas in the latter no differences
in the ionization fraction were detected between cores with and 
without associated stars.  No changes in the ionization fraction was found 
either between the edge and the center of the mapped 
regions.  However, no similar studies have been performed so far in the 
direction of pre--star--forming cores in relatively quiescent zones of our 
Galaxy, which give the opportunity to investigate the initial conditions
in the star formation process presumably not triggered by external agents.

 In any model of a pre--protostellar cores, one expects to observe a 
 concentration of material around the region of highest density. As the 
 density becomes higher, one expects the degree of ionization to decrease,
the degree of depletion onto dust grain surfaces to increase, and, 
consequently, the fractionation of deuterated species to be enhanced 
(e.g. ~\cite{RM00}). In fact, one might naively
 expect that deuterated species correlated with high depletion 
might trace the phase
 prior to the formation of a protostar. With this in mind, we
 undertook observations using the IRAM 30-m telescope of a variety
 of species towards the dense core L1544 in the Taurus molecular cloud
which is thought (e.g. ~\cite{TMM98}, ~\cite{CB00}) to be in a phase shortly 
prior to that at which collapse occurs.  The data and a kinematical study
of this region are presented in Caselli et al. (2001, hereafter Paper I).
Here we present our determination of the electron fraction across L1544
and show that $x(e)$ at the center of the core -- where the H$_2$ number
density is $\sim$ 10$^{6}$ cm$^{-3}$ -- is about 2$\times$10$^{-9}$,
about five times smaller than expected if the electron fraction is due 
to cosmic--ray ionization alone and the cosmic--ray ionization rate 
$\zeta$ has its ``standard'' value ($x(e)$ = 1.3$\times 10^{-5} 
n({\rm H_2})^{-0.5}$; cf.~\cite{M89}).  

 In section 2, we show our determination of physical and chemical 
parameters of importance for electron fraction estimates. 
The determination of electron fraction is presented in Section 3.  
Finally in Section 4, we discuss the consequences of our
 measurements for the understanding of high density cores like
 L1544.

\section{Determination of physical and chemical parameters}

An accurate estimate of electron fraction implies measurements
of (i) the column density of the observed molecular ions, (ii) the temperature 
as well as (iii) the density structure of the core, and (iv) 
the amount of CO depletion.  In the following sections we will 
describe the above points separately. 

\subsection{The temperature across the core}
\label{stemperature}

Because of its low dipole moment, CO is likely thermalized at the
typical core densities and it can be used to measure kinetic temperatures.
~\cite{TMM98} used the J = 1$\rightarrow$0 transition of
\TWCO, \THCO ~and \CEIO ~to determine
the excitation temperature of the CO emitting gas across L1544 and found
$T_{\rm ex}$ $\sim$ 12 K, in good agreement with previous kinetic temperature
estimates based on CO and \AMM ~observations ($\sim$ 10 K; \cite{MLB83},
~\cite{UWW82}, \cite{BM89}).
We repeated a similar analysis using \CEIO (1--0) and \CSEO (1--0) data
and found an average value of $T_{\rm ex}$ = 10.1\p0.1 K .
Of course, CO is not a good tracer of gas temperature
at densities $\ga$ 10$^5$ \percc , where it is strongly depleted (CWT99;
see Sect.~\ref{sdepletion}).
A recent Monte Carlo radiative transfer
analysis of the two inversion transitions of \AMM, observed at the
Effelsberg antenna (Tafalla et al., in prep.; 40$^{\prime\prime}$ half power
beamwidth), has also shown that
$T_{\rm k}$ $\sim$ 10 K with no significant variation across L1544.  We
assume on this basis 
a constant temperature of 10 K in the rest of our analysis.
However, model calculations of the dust temperature in pre--stellar cores
(~\cite{ZWG01},~\cite{ERSM01} ) predict a 
fall off in dust temperature from $\sim$ 14 K at the edge to $\sim$ 7 K at  
the center.  This is likely to be reflected also in the behaviour of the
gas temperature at least for densities above 10$^5$ cm$^{-3}$. 
Although for our purposes we can 
neglect such findings (our simple chemical model presented in 
Sect.~\ref{smodel} is not significantly affected by a 
change in temperature from 10 to 7 K), the predicted temperature gradient  
affects our 
estimates of the volume density from \NTHP , as discussed below.

\subsection{The density across the core}
\label{sdensity}

From our molecular line data we can deduce good number density
estimates only at the L1544 ``molecular peak'' and in a few adjacent
positions (see Sect.~\ref{sdcod}, \ref{sn2dpd}).  
Therefore, for our electron fraction estimates across
L1544 (Sect.~\ref{smodel}) we used the density profiles obtained from
 1.3mm continuum observations (\cite{WMA99}, hereafter WMA) 
and ISOCAM absorption in the 
mid--infrared (\cite{BAP00}, hereafter BAP).  
These density profiles are described in Section~\ref{sdescription}. 

A number of line intensity ratios measured by us are sensitive to density.
 Since the line ratios depend on collisional rates, one in principle
 derives n(\MOLH) independently of assumptions about abundance.
 However, in a situation where the density varies rapidly along
 the line of sight (as suggested by the dust emission and absorption),
 the density which one derives will be a weighted mean. In the following,
 we describe some estimates we have made using different tracers.

\subsubsection{Density inferred from \DCOP (3-2)/\DCOP (2-1)}
\label{sdcod}

 Our observations of  \DCOP (3--2) and (2--1) allow estimates to be made
 of the hydrogen number density in the regions responsible for the emission of
 these lines as well as estimates of the \DCOP \ column density. We have
 done this using an LVG statistical equilibrium code similar
 to that described e.g. in ~\cite{W87}. The approach is essentially identical
 to that discussed by ~\cite{BLL95}. The collisional rates were
 taken from those calculated for collisions of para-H$_2$ with \HCOP \ by
 ~\cite{M85} assuming a temperature of 10 K. At this temperature,
 the rates are essentially identical to the more recent rates of
 ~\cite{F99}. We assumed a dipole moment of 3.9 Debye following
 ~\cite{B89} (see also ~\cite{WBC98}).

 In Fig. ~\ref{fnh2}, we show the results obtained in this manner
 superimposed upon the half--maximum contours of the \DCOP (2-1) and
 (3-2) maps (see Paper I for details on these data) 
and the 1.3mm continuum map of WMA. We see a definite
 tendency  for the highest densities derived from \DCOP \ to be
 close to the dust emission maximum. We see also that the values found
 for n(\MOLH) towards the dust peak
 are close to those derived by WMA from their observations ($\sim$ 10$^6$
cm$^{-3}$). However,
 there are a number of {\it caveats} of which one should be aware.

 One of these is that the \DCOP (2-1) line is almost certainly optically
 thick towards the core of L1544. This is suggested by
 both our LVG results themselves and the comparison at offset (20,-20)
 with \DTHCOP \  (see Paper I) 
which suggests an optical depth of order 4.  Our density
 estimates are thus sensitive to the geometry and velocity gradients
 within L1544.  We note additionally that our \DCOP (3-2) map  cannot 
be convolved to the 2--1 resolution (1.5 times worse) because it 
is undersampled  (grid of 1.7 times HPBW).  Nevertheless, our results 
suggest that we are sampling in \DCOP \ the high density inner core of L1544.

\subsubsection{Density derived from \NTHP and \NTDP}
\label{sn2dpd}

 Our \NTHP \ and \NTDP \ observations allow an independent estimate
 of the density. This is of special interest because, as we have seen in 
Paper I,
 the \NTHP \ and, in particular, the \NTDP \ integrated intensity
 maps appear to trace best the dust emission (see also \cite{BCL01}).  
Another characteristic
 of these species is that, due to the hyperfine splitting, we can
 determine directly the optical depth in several transitions utilising
 the relative intensities of the hyperfine satellites.  This
  reduces considerably the errors in our determinations of level
  column densities for, say, \NTHP \ , as compared to \DCOP .
  Finally, it is of some importance that we have detected \NTHP (3-2)
  and \NTDP (3-2). This is because we expect such transitions to be
  essentially collision dominated in the  sense that the integrated
  line intensity has, for a given temperature, a one--to--one
  correspondence with the integral of the product of the \MOLH \
  density and the \NTHP \ density along the line of sight.
  We call this integral $\int n(\MOLH ) n(\NTHP) ds$ the
  \NTHP \ collision measure $CM(\NTHP)$ in the following.
  Essentially, the (3-2) line intensities are proportional to
  $CM(\NTHP)$ in the limit where collisional
  deexcitation is negligible with respect to radiative transitions
  which for \NTHP (3-2) at 10~K for example implies n(\MOLH ) less
  than $10^6$ \percc .

  We have developed an LVG code for \NTHP (\NTDP ) analogous to
  that for \DCOP . We use the collisional rates of Green(1975)
  for collisions of \NTHP \ with He and multiply them by 1.4 to
  take account of the differing mass of \MOLH .  We also have
  adjusted the  escape probability to account for hyperfine
  splitting in the 1-0, 2-1, and 3-2 transitions of \NTHP .
We first use our determination of
  the optical depth in the \NTHP (1-0) line at offset (20, -20) 
to obtain an estimate  of the \NTHP \ column density and find
  $N(\NTHP )$ = $1.5\, 10^{13}$ \cmsq \ at offset (20, -20). Then,
we determine $CM(\NTHP )$ from \NTHP (3--2) data, which have been 
averaged in a 3$\times$3 grid of positions spaced by 10\arcsec \ and
centered at (20, -20) (see Tab.~3 of Paper I, second row) 
in order to simulate a 20\arcsec \ beam, similar
to the 1--0 observations.
From the $CM(\NTHP )/N(\NTHP )$ ratio we find a mean density of
1.3$\times$10$^5$ \percc , smaller than the value derived above from our
 \DCOP \ measurements and also smaller than the central densities
 suggested by the results of WMA and of BAP.
These discrepancies may be partly due to the 
assumption of a constant temperature of 10 K, 
in particular in the dense highly
CO--depleted nucleus of L1544, best traced by N--bearing molecules, where
\cite{ZWG01} derive $T$ $\sim$ 7 K.

%
%


 We can in analogous fashion use our \NTDP \ observations to obtain
 a density estimate. At the center of L1544 assuming a temperature of
 10~K, we find a mean density 
$CM(\NTDP)/N(\NTDP )$ = $1.5\, 10^{18}/4.9\, 10^{12}$ =
 $3\, 10^5$ \percc .  This is a factor of 2
 higher than the \NTHP \ estimate above which might perhaps reflect
 the fact that the \NTDP \ emission is more compact than that of
 \NTHP .  However, it is still a factor of 3 smaller than the density
deduced from dust continuum and \DCOP \ observations, again suggesting
a possible temperature fall off in the core center.

\subsection{Depletion of CO}
\label{sdepletion}

The depletion factor $f_{\rm D}$ is defined by the ratio between the
``canonical'' fractional abundance of CO (9.5$\times$10$^{-5}$, \cite{FLW82}
\footnote{We note that the only direct measurements of the 
CO abundance give larger values  than those reported by 
\cite{FLW82} by a factor of about 5 (\cite{LKG94}), implying  
larger depletion factors than found in this work.})
and the observed $x$(CO) = $N(\rm CO)/N(\rm H_2)$.  To determine $f_{\rm D}$
at the dust peak position, CWT99 compared the \CSEO ~integrated
intensity with the 1.3mm continuum dust emission flux density from WMA.
They found that $f_{\rm D}$ $\sim$ 10 at the
dust peak, and that observations are consistent with a model where CO is
condensed out onto dust grains at densities $\sim$ 10$^5$ cm$^{-3}$.  The
corresponding radius of the region where CO is severely depleted is $\sim$ 6500
AU and the depletion causes 2.3 \solmass \ of gas to be lost to view in 
molecular line emission.

Here we extend this analysis to all the other positions where \CSEO ~has
been observed.  To do this we divided the 1.3mm map by the \CSEO (1--0) map.
This division implies three steps: (i) convolve the 1.3mm map with a 22\arcsec \
beam (the HPBW of the \CSEO ~observations) and reproject it to have the same
coordinates as the \CSEO ~map; (ii) make a regularly sampled \CSEO ~map;
(iii) divide one map by the other.  The result of this operation can be
translated into a depletion factor map.  
We somewhat arbitrarily assume ``normalcy''
to be given by [\CSEO ]/[H$_2$] =
4.8$\times$10$^{-8}$ (\cite{FLW82}).  Assuming an excitation temperature
for \CSEO \ of 10 K  and a 1.3 mm dust grain opacity
 $\kappa_{\nu}$ = 0.005 cm$^{2}$
g$^{-1}$ (WMA) , we find that the
 depletion factor $f_{\rm D}$ can be expressed as:
\begin{eqnarray}
f_{\rm D} & = & 0.027 \frac{S(1.3{\rm mm, mJy/22 \arcsec})}
 {W_{\CSEO}({\rm K} \, \kms)}.
\label{efd}
\end{eqnarray}
 Here, $W_{\CSEO}$ is the integrated intensity ( over all hyperfine
 components) of \CSEO (1-0) and $S(1.3{\rm mm, mJy/22 \arcsec})$ is
 the flux density at 1.3mm in a 22\arcsec \ beam.  In those positions
where only \CEIO \ was observed, we used $f_{\rm D}$ =  0.085 $S(1.3{\rm mm, 
mJy/22 \arcsec})/W_{\CEIO }({\rm K} \, \kms)$.

In Fig.~\ref{fdep} the depletion
factor map is shown together with the 1.3mm continuum map from WMA.
There is a fairly good correspondence between the $f_{\rm D}$ and
1.3mm contours as one might expect given the ``flat'' nature of the
\CSEO \ and \CEIO \ maps.  We note however that the peaks are
not coincident and there is a  projected distance between the
$f_{\rm D}$ and the dust peaks is 13\arcsec ~(about half a beam).
This appears significant but one must remember that the \CEIO \
and \CSEO \ maps are probably dominated by emission from gas of
density a few times $10^4$ \percc \ to the foreground or background
of the dust emission core.  Thus structures unrelated to the
pre--protostellar core may influence our maps.

\subsection{Molecular ion column densities}
\label{sder}

The methods used to estimate column densities of the 
observed species, which give same results within a factor
of 2, are reported in appendix~\ref{acolumn}.  Here we 
show the results. 

\subsubsection{\DCOP \, and \HCOP }
\label{sdco}

The statistical equilibrium
calculations used
to estimate the \MOLH \ density (see Sect.~\ref{sdensity}) 
can also be applied to a determination
of the column densities of \DCOP \ and \HCOP .  As in the case of
density determination, our errors grow rapidly if the observed lines
are optically thick.  Towards the peak of L1544, we  have
used the optically thin species \HCEIOP \ and \DTHCOP \ 
to determine \HCOP \ and \DCOP \ column densities 
and \CSEO \ to determine the CO column density. We have
considerable confidence that in these three cases, the observed
lines are in fact optically thin (see discussion in Sect.~\ref{adcop}).
 For \CSEO , we can check (see e.g. CWT99 )
optical depth using the relative intensities of the hyperfine
satellites.

 We nevertheless for most of the following discussion use \HTHCOP \
 and \DCOP \ column densities derived from transitions
 which are optically thick towards the dust emission peak and in
 the immediate vicinity. The reason for this is the greater extent of
 our maps in these lines. The justification is that our
 checks with the optically thin variants yield results which are
 consistent with those presented here.  Thus we have checked our
 \HTHCOP \ column densities using \HCEIOP \ where available and
 our \DCOP \ column density using the \DTHCOP \ measurement at
 (20,-20).  We conclude that our column density estimates
 are accurate to within a factor of 2 also when the optical depth is
 high (see appendix~\ref{adcop} for details).

 Bearing this in mind, we present in Fig. ~\ref{dco-hco}
 our  \DCOP \ and \HCOP \ column density maps.  One sees in the
 first place that these differ considerably from the integrated
 intensity maps. While for the reasons discussed above, the
 column density maps (or ratio maps)
 should be treated with caution, we  consider them a better
 estimate than  simply scaling the integrated intensity maps.
 We also show in Fig. ~\ref{dco-hco} the inferred ratio of column
 densities $N(\DCOP)/N(\HCOP)$.  \DCOP \ is clearly more concentrated
 towards the structure seen in dust emission by WMA than is
 \HTHCOP . Considering the column density ratio map, we
 see that [\DCOP]/[\HCOP] appears to vary by almost an order of
 magnitude between the centre and the edge of the map (roughly
 an arc minute or 0.04 parsec).  
 Away from the peak, the inferred $N(\DCOP)/N(\HCOP)$ ratios are
 considerably lower than found in the survey of ~\cite{BLL95}.
  This is somewhat surprising because we expect to be
 observing higher density gas on average with greater degrees of
 depletion than in the arc minute resolution
 \cite{BLL95} survey.  The ~\cite{BLL95} estimates for
 $N(\DCOP)/N(\HCOP)$ vary between 0.02 and 0.07 whereas our results for
 L1544 vary from 0.04 near the dust peak to values a factor of
 10 lower at the edge of our map.

\subsubsection{\NTDP \, and \NTHP}
\label{sn2hp}

The \NTDP \ and \NTHP column density maps, together with 
$N$(\NTDP )/$N$(\NTHP), are also shown in Fig.~\ref{dco-hco}.  
 Our \NTDP \ observations as noted previously (see also Paper I) show that
 \NTDP \ {\it traces the dust peak emission better than any other species}.
 It also turns out that the values we infer for $N(\NTDP )/N(\NTHP )$
 are much higher than for $N(\DCOP )/N(\HCOP )$. At offset (20,-20), we infer
 $N(\NTDP )/N(\NTHP )$= 0.24$\pm 0.02$.  This is a factor of
 4 higher than our maximum value for $N(\DCOP )/N(\HCOP )$.  We conclude from
 this that the degree of deuterium enhancement increases towards the
 dust emission peak of L1544.

\section{The electron fraction $x(e)$}
\label{sxe}

CWT98 used a simple chemical model to find 
analytic expressions which allow one to directly estimate $x(e)$ (and 
the cosmic ray ionization rate $\zeta$) once $N(\DCOP )/N(\HCOP )$,
$f_{\rm D}$, and $n({\rm H_2})$ are known from observations (see their 
eqns.~3 and 4). However, these expressions furnish upper limits
of $x(e)$ and $\zeta$ (see Section \ref{sul}).  In fact, in the simple 
chemical scheme used to derive CWT98 analytic expressions, dissociative 
recombination of molecular
ions onto negatively charged dust grains and gaseous reactions
involving atomic deuterium (see e.g. ~\cite{DL84}) are not taken 
into account.  Moreover, this chemical model is only valid for homogeneous 
clouds, whereas L1544 clearly has a density structure and a varying amount of 
depletion and molecular abundances along the line of sight (see 
Sect.~\ref{sder}).  In fact, CWT98 
eqns.(3) and (4) have been applied to data with a significantly lower (factor 
of $\sim$ 6) spatial resolution compared to the present observations, so that
any gradient was smoothed out in the beam and resultant $N(\DCOP )/N(\HCOP )$, 
$f_{\rm D}$, and $n({\rm H_2})$ should be considered as ``average'' quantities
for the cores.  The same problem is met if ``pseudo--time dependent'' 
chemical codes are used.   The density structure as well as differential 
depletion of molecular species has to be taken into account to reach a 
satisfactory agreement with observations (see also \cite{AOI01}).  

In this section, we discuss what one can say about the ionization
fraction in L1544. One may obtain a lower limit to the ionization
fraction $x(e)$ summing the abundances of observed
molecular ions in L1544. This limit may  be close to the real
ionization fraction, if refractory metals of low ionization potential
such as Mg and Fe are significantly depleted (see e.g. 
Fig.~2 of \cite{CWT98}).

We first briefly discuss the limits one can place on ionization degree
based upon the observed molecular ion column densities.

\subsection{Lower limits for $x(e)$}
\label{sul}

Lower limits of the electron fraction across L1544 can be
obtained by simply summing up column densities of the observed
molecular ions:
\begin{eqnarray}
x(e)_{l} & \geq & \frac{N(\HCOP )+N(\DCOP )+N(\NTHP )+N(\NTDP )}{N(\rm H_2)}
	\sim \frac{N(\HCOP )}{N(\rm H_2)}.
\end{eqnarray}
$x(e)_{l}$ has been obtained for each observed position toward L1544
based upon the column density determinations from Sect.~\ref{sder} and the
column densities of WMA.  We find in this way
a lower limit at essentially all positions where \HTHCOP \ was
observed of :
$<x(e)_l>$ $\sim$ 1$\times$10$^{-9}$ with little variation.

 These estimates do not however really sample the
 high density nucleus of L1544 and in order to do that, we use the
 model discussed in the following section.

\subsection{$x(e)$ from chemical models}
\label{smodel}

Deriving the behavior of the electron fraction as a
function of radius is complicated however by the strong
 abundance gradients  due to depletion. These cause observed
 quantities such as e.g. the \HCOP \ column density to be strongly
 influenced by background and foreground emission in the observed
 lines. This is the
main rationale for the simple model introduced below.

\subsubsection{Model description}
\label{sdescription}

    The results of WMA and BAP
 show rather clearly that the L1544 ``core'' itself contains a
 high density ``core''  where the density reaches values as large
 as $5\, 10^5$ \percc \ over a volume of radius several thousand
 AU. We have shown previously (CWT99) that CO is
 depleted  by at least an order of magnitude in the central
 high density peak. In this section, we consider the behavior of the
 ions and of the electron density in the depleted region.   We
 have done this with an extension of the model briefly described by
 CWT99 and used to fit the CO depletion.

 We assume for this purpose a spherically symmetric
 density distribution with a constant molecular
 hydrogen density of (i) $5.4\times10^5$
 \percc \ out to  a radius $r_{\rm flat}$ = 2900 AU followed by a
 $1/r^2$ fall off out to a cut--off radius of 10000 AU, which
 simulates the results of BAP based on
 the extinction measured with ISOCAM against the emission of small
particles at a wavelength of 7 $\mu$m; (ii) 1.3$\times$10$^6$ \percc \
out to a radius $r_{\rm flat}$  = 2500 AU followed by a $1/r^2$
fall off out to a cut--off radius of 13000 AU, as found by WMA on the basis
of 1.3mm observations of dust emission.
The H$_2$ column and volume density inside $r_{flat}$ determined by BAP
are a factor of 3.8 and 2.4, respectively,  lower than those found by WMA
(see however the comments in Table 2 of BAP).
It is clear that this approximation is somewhat
 crude and indeed that the distribution departs considerably from
 spherical symmetry. Moreover, the recent study of \cite{ERSM01}
has shown that the temperature drop in the L1544 core nucleus to about 
7 K (in agreement with \cite{ZWG01}) implies a steeper density profile 
than in the isothermal case of WMA and BAP.
  However, we believe that our simple model
  suffices to explain many of the observations.

  As in CWT99, we have followed the time--dependent depletion
  of CO neglecting dynamic effects. That is to say, we assume the
  depletion time--scale to be more rapid than dynamical
  time--scales and assume initially a ``canonical'' CO abundance
  of $9.5\times10^{-5}$ relative to \MOLH . However, we  also
  consider in this study the recycling of CO back into the
  gas--phase using the cosmic ray desorption rate proposed by
  ~\cite{HH93} and a CO adsorption energy $E_{D}$(CO) to the
  grain surface of 1210 K (\cite{HHL92}).   This is for a surface
  of SiO$_{2}$  while for binding with a CO mantle, Sandford
  and Allamandola (1990) find $E_{D}$(CO) to be 960 K.
  This is critical for the model because
  we have, following ~\cite{HH93}, assumed a time--scale
  for cosmic ray desorption $t_{cr}$ of species $X$ given by :
  \begin{equation}
   1/t_{cr}(X) \, = \,  3\, 10^{-7}\, \zeta _{17} \times \exp [{-E_{D}(X)/70}]
   \end{equation}
   Here, $\zeta_{17}$ is the cosmic ray ionization rate in units
   of $10^{-17}$ s$^{-1}$ and $E_{D}(X)$ is the adsorption
   energy in K of species $X$.
  Rough estimates 
  suggest that ``chemical'' desorption due to molecular hydrogen
  formation (\cite{WM98}) appears to be less effective.  Indeed, 
a recent theoretical work by  ~\cite{TW00} has shown that the chemical 
desorption of CO can occur on small grains with size less than 20
$\AA$, but it is negligible on larger grains.  
  A crucial point in our discussion moreover is that we also
  consider the analogous depletion of molecular nitrogen assuming an
  initial \MOLN \ fraction of 7.5$\times$10$^{-5}$ (\cite{MCS97}).
  This is a maximal value in that it is based on nitrogen depletion
  measurements in diffuse clouds. Critical for our model is
  also that we adopt a \MOLN \, adsorption
  energy of $\leq$ 787 K (= 0.65$\times E_{\rm D}$(CO);
  \cite{BL97} based upon calculations of ~\cite{SRD95})
  considerably
  lower than that for CO. The exponential dependence of the cosmic
  ray desorption rate above causes more effective depletion of CO
  and gives rise to a layer where \MOLN \ is the most abundant
  gas phase species containing heavy elements and where \NTHP \ is an
  abundant ion.

  In a model of this type, one can write for the abundance $n(X)$ of
  species $X$ (where $X$ can be CO or \MOLN ) relative to \MOLH \  that :
  \begin{equation}
  n(X) \, = \, n(X,\infty)\, + \, (n_{tot}(X)-n(X,\infty))\, \exp{-(t/t_{0})} .
  \end{equation}
  Here, $n_{tot}(X)$ is the total abundance of $X$ in both gas and
  solid phase , $n(X,\infty )$ is the steady state
  gas phase abundance given by
  $n_{tot}(X)/(1\, + t_{cr}/t_{dep})$  and $t_{dep}$ is the depletion time
  scale $1/(S \, n_{gr}\sigma _{gr}v(X))$  (where $S$ is the sticking coefficient,
  $n_{gr}$ is the grain number density, $v(X)$ is the thermal velocity of $X$,
  and $\sigma _{gr}$ is the grain cross section).  The time--scale $t_{0}$ is
  simply $(t_{dep}\, t_{cr})/(t_{dep}+t_{cr})$.  We are assuming here
  that one can neglect destruction of $X$ due to either gas or
  solid--phase reactions.

 This rather crude model of depletion has the advantage that it can be rapidly
 computed as a function of density and time. For the present purpose, we
 (as in CWT99) continue the calculation until such time as
 we reach a central CO column density compatible with the observed \CSEO
 \ column density toward the central dust peak in L1544. At this point, as
 discussed by CWT99, the main contributions to the observed CO
 column come from lower density foreground and background material while
 CO in the central high density peak is depleted to very low abundances.
 The present calculations confirm this.
Even though cosmic ray desorption is included, the central CO abundance 
is still depleted by a factor of 10$^3$ ($\sim$ 30), if the WMA (BAP) density 
structure is used.  This
 also depends sensitively on $E_{D}(\rm CO)$ and indeed
 our results suggest that $E_{D}(\rm CO)$
 must be larger than $\sim$ 900 K in order to
 have depletion in the central region as observed (for the standard values
of $\zeta$ and $S$, see Tab.~\ref{tbfit}).  A similar result has also been 
found by ~\cite{AOI01}, who cannot reproduce the observed ``hole structure''
in the CO column density if the grain surface is covered by 
non--polar ice (i.e. $E_{D}(\rm CO)$ = 960 K).

 Although the gas phase chemistry in general has long time scales, the
 ion chemistry is relatively rapid and thus one can expect in
 reasonable approximation that the abundances of ionic species such
 as \HCOP , \NTHP \ etc. are given by the steady state chemical
 equations using the instantaneous abundances of the neutral species
 which are important heavy element repositories.  The time scale
 for the ``ion chemistry'' is expected to be determined by
 recombination to species such as \HCOP \ with a rate of
 $\beta _{diss}\, n(e)$ where $\beta _{diss}$ is the
 dissociative
 recombination rate and $n(e)$ the electron density. For
 ``canonical
 estimates'' $n(e)\sim 10^{-3}$ \percc \ and $\beta _{diss}\sim
 10^{-6}$ cm$^{3}$ s$^{-1}$, we find a time scale of
 30 years for the ion chemistry which is much less
 than depletion time scales. Thus one can compute
 \HCOP \ in terms of the instantaneous CO abundance
 and \NTHP \ in terms
 of the instantaneous \MOLN \ abundance.

 Analogously, the electron
 fraction $x(e)$  can be computed in terms of global estimates for the
 molecular and metallic ions and using the  same instantaneous
 abundances for CO, \MOLN \ etc. in the gas phase.
 The approach we have adopted is a simplified version
 of the reaction scheme of ~\cite{UN90}.
 Thus, we compute a
 generic abundance of molecular ions ``mH$^{+}$''  assuming formation
 due to proton transfer with \HTHP \ and destruction by dissociative
 recombination with electrons and recombination on grain surfaces
 (using rates from ~\cite{DS87}).  The abundance of  \HTHP \
 is calculated considering formation as a consequence of cosmic ray
 ionization of molecular hydrogen and destruction by proton
 transfer with heavy molecules (essentially CO and \MOLN \
 here) as well as due to the processes mentioned above for molecular ions.
 We also consider metal ions M$^{+}$ formed by charge transfer  and
 (including the equation for charge neutrality) solve a cubic
 equation for the electron abundance.

  In similar fashion, we have computed the instantaneous
  [\DCOP]/[\HCOP] (and [\NTDP]/[\NTHP]) ratios assuming that the
  respective deuterated species form by proton transfer from
  \HTDP .  Thus,  we express  the abundance ratio 
	$R_{D} = [\DCOP]/[\HCOP]$ as :
  \begin{equation}
 R_{D} \, = \, (1/3)\, k_{1}\, x(HD) \, /
 \, (k_{e}x(e)\, + \, k_{2}x(m)\, +\, k_{3}\, x(gr)).
 \label{edeut}
  \end{equation}
In the above, $k_{1}$ is the rate for production of \HTDP \
due to reaction of HD with \HTHP , $k_{e}$ is the rate for
  dissociative recombination of \HTDP ,  $k_{2}$ is the 
rate coefficient for \HTHP \ (and \HTDP ) destruction with neutral 
species, and $k_{3}$ is the
  rate for recombination of \HTDP \ on grain surfaces (\cite{DS87}).
$x(\rm HD)$ is the HD abundance relative to \MOLH \
  taken to be $3\times 10^{-5}$ (\cite{LDW95}), $x(m)$, a function 
of the amount of depletion,  is
  a weighted average over neutral molecules which can undergo
  proton transfer with \HTHP , and $x(gr)$ is the grain abundance
  by number.  The latter has been computed
  (as has $k_{3}$) using a MRN (\cite{MRN77})
  distribution with lower cut--off radius a$_{min}$
  (100 \AA \ in standard case) and
  upper cut--off 0.25 $\mu$m.  This treatment ignores effects due
  to atomic deuterium (\cite{DL84})  and moreover
    assumes that all deuterium enrichment originates in
    \HTDP \ (rather than e.g. CH$_2$D$^+$). However, it improves
 on equation (1) of CWT98 in that it treats the
 depletion (via $x(m)$) in consistent fashion and in that it
 considers the effect of recombination on grains. From eqn.~\ref{edeut}
it is clear that an increase of depletion boosts the  
deuterium fractionation.  In fact, depletion causes a drop in the 
\HTDP \ and \HTHP \ destruction rates, and a rise in the \HTDP \ 
formation rate, via \HTHP + HD, due to the larger \HTHP \ abundance. The net 
result is a larger \HTDP /\HTHP \ abundance ratio and a consequently 
more efficient deuterium fractionation.

 In Fig.~\ref{fstandard}, we show the dependence on radius
 of various ionic and molecular abundances predicted by our
 standard models fitting the central CO column density of
 L1544 (left panels assume the density structure fit to the
 results of WMA,
whereas the right hand panels were  obtained assuming the density
structure deduced by BAP).  The top panels show predicted depletion factors for
 CO and \MOLN . The bottom panels shows the electron fraction,
 and the predicted abundances for the various ionic species
 observed by us. We stress that this is {\it not} our
 ``best fit'' model (see Sect.~\ref{sbestfit}) but it illustrates several
 features of our results.

 We note first the rapid increase of the CO depletion factor with decreasing
 radius, i.e. the gaseous
CO abundance is reduced to a value of order 10$^{-3}$
(0.03) of the ``canonical value'' of $10^{-4}$ relative to \MOLH \ in
 the constant density central region, using WMA (BAP) density values.
 This contrasts with N$_2$ which (although itself depleted) becomes
 the most abundant heavy molecule in the center.  It is the rapid variation of
 these quantities as a function of density which renders necessary
 a consideration of abundance variations along the line of
 sight and hence radially.
 Also shown in Fig.~\ref{fstandard} are the variations of the \DCOP \
 and \HCOP \ abundances. Their ratio ($R_{D}$)
 increases by roughly an order of magnitude from 0.04 (0.04) to 0.7 (0.35),
using WMA (BAP) parameters, between the edge and center of our model.

 The variation in ion densities shown in the bottom panel of
 Fig.~\ref{fstandard} reflects the  behavior of the depleted species
in the two standard models.  Thus, in a WMA core, \HCOP \ abundance
is strongly reduced in the center because of
CO depletion, whereas \DCOP \ shows a rather flat dependence due to the
increase of the deuterium fractionation with increasing $f_{\rm D}$.
A similar behaviour is present for the \NTHP \ and \NTDP \ abundances,
with the consequence that $R_{\rm D}$ and [\NTDP ]/[\NTHP ] become
greater than 1 in the innermost parts of the core.  In a BAP core,
the lower density causes
less efficient CO and N$_2$ depletion, and
much shallower profiles for \HCOP \  and
\NTHP \ fractional abundances. The ions plotted in the figure are ``major
 ions'' at all radii (except that \HTHP \ becomes abundant in the
 central region under some circumstances) suggesting that
 for depleted high density cores, the sum of the observed
 molecular ion abundances may give a good estimate of the
 electron abundance.  This is a great simplification
 relative to ``normal cores'' where ``metal ions'' have
 comparatively high abundances. Our calculations
 suggest that in high density
 depleted cores, the depletion of ``metals''  (assumed here to
 behave like CO) is likely to be
 so large that this complication becomes negligible.  We have used 
an initial metal abundance about one order of magnitude lower than 
that measured in diffuse clouds,  as is usual (``low metals'') 
in chemical models of dense clouds (e.g. \cite{PH82}; 
\cite{HL89}; \cite{GLF82}; \cite{LBH96}).

 The most convincing aspect of our model is that it
 qualitatively explains the enhancement of \NTDP \
 in the center of L1544, close to the  depletion peak. Thus 
it explains the higher degree of fractionation of deuterium 
in \NTDP \ as compared with \HCOP .  It also explains roughly 
the differing depletions observed for \NTHP , CO, and \HCOP . 
 In fact, as discussed
 in the next section, the qualitative features of the model are in excellent
 agreement with observation. The pattern that ions
 and deuterated species peak close to dust emission maxima appears to
 be a general feature (\cite{TMC01}) of nearby cores and
 suggests that the model presented here has wider applications
 than to L1544.

\subsubsection{The best fit model}
\label{sbestfit}

Several model calculations have been performed to find the
``best fit model'', i.e. the model which best reproduces observed
column densities at the peak of L1544.  
We started with a ``standard'' cloud model,
where $\zeta$ = 1.3$\times$10$^{-17}$ s$^{-1}$, the CO binding
energy $E_{\rm D}$(CO) = 1210 K (\cite{HHL92}), the N$_2$ binding
energy $E_{\rm D}$(N$_2$) = 787 K, the minimum radius of dust grains
$a_{\min}$ = 10$^{-6}$ cm, the sticking coefficient $S$ = 1.0
(see Table~\ref{tbfit}).
Two density distributions (from WMA and BAP) were considered
and the above parameters were varied  until the quantity:
\begin{eqnarray}
\chi^2 & = & \sum_{i=1}^4 \left[ \frac{1}{\sigma_{N_{\rm obs}(i)}} \left(
	N_{\rm obs}(i) - N_{\rm mod}(i) \right) \right]^2 \label{echi}
\end{eqnarray}
was minimised.  In eqn.(\ref{echi}), $N_{\rm obs}(i)$ and
$N_{\rm mod}(i)$ are the observed and model calculated column density
of \HCOP , \DCOP , \NTHP , and \NTDP \, at the L1544 ``molecular
peak''.  $\sigma_{N_{\rm obs}(i)}$ is the uncertainty
associated with $ N_{\rm obs}(i)$, which has been assumed here
to be a 30\% error. Thus we assume the errors in our column density
determination to be dominated by systematic effects such as those
caused by our summary treatment of the radiation transfer.

The column density dependence as a function of impact
parameter $b$ predicted by the two ``best fit'' models
(for the two density distributions)
can then be compared with those  observed. We have made azimuthal
averages of our observed column densities
 using averages within bins defined by
 $i < b < i+20$ arcsec, with $i$ = 0, 20, 40, ... (arcsec),
with the exception of the value at $b$ = 0.  This is shown in
Fig.~\ref{fmol_dist}, where large symbols represent the average column
density inside the corresponding bin (symbols mark the upper edge of each
bin).  One sees that there is considerable scatter as indeed one
can expect given the elongated nature of the maps. This is
in particular  the case for \NTHP \ where the ``mean'' calculated in
this fashion may clearly be misleading. One sees nevertheless
that the distributions become clearly more peaked going from CO
(flat) to \NTDP .

In Fig.~\ref{fbfit},
the observed column density profiles are compared with
those predicted by four  models compatible with the
peak column densities.  The parameters used
in these models are reported in Tab.~\ref{tbfit} together with the
$\chi $--squared estimates using equation (\ref{echi}).   To
understand our results, it is useful to recall that we need to
assume high degrees of CO depletion in order to explain the CO and \HCOP \
column densities.  A certain amount of nitrogen depletion is also
needed to explain N(\NTHP ).  Such low gas phase molecular abundances
cause extremely high degrees of D fractionation however with the
model expectation for e.g. [\DCOP]/[\HCOP] higher than unity in the
dense depleted central regions. This then gives difficulty explaining
the abundances of \NTDP \ and \NTHP .  The range of parameters in
Tab.~\ref{tbfit}  demonstrates that the peak column densities on
their own only give partial constraints on the chemical models.

 Fig. ~\ref{fbfit} however shows that the extent
 of the emission in the various ions is also an important constraint.
 In models 1 and 2 (Tab.~\ref{tbfit}) for example, the model
 half--width for \NTHP \ is much too small. 
 While this could be
 partially due to our azimuthal averaging (see Fig.~\ref{fmol_dist}), 
it is also the case that the cosmic ray ionization rate has been chosen
 to have values in models 1 and 2 more than an order of magnitude
 lower than in the standard model
  and the result is that N(\NTHP) is much too small at
 large radii.

A way of improving the agreement between observations and model predictions
is to more efficiently deplete CO outside the core center (to increase the
\NTHP \ column density at $b$ $>$ 0).   This can be done by increasing
the CO binding energy and thus reducing the efficiency of cosmic ray
induced desorption. However, this also causes an
increase in  deuterium fractionation which is not observed.  As
discussed in Sect.~\ref{sdco}, [\DCOP]/[\HCOP]  is observed to have rather
low values at high impact parameter.
This suggests that reactions other than with
CO and \MOLN \ act to reduce the \HTDP \ abundance.
We therefore postulate the existence of
a volatile neutral species  which will remain in the gas phase when
CO is depleted and which has a proton affinity permitting transfer
of a proton from \HTHP .

  This species needs to be abundant (at least as abundant
as CO) to effectively keep the deuterium fractionation low in those
regions where CO is significantly depleted.  One possibility may be
atomic oxygen, which is predicted to be quite abundant in dense clouds
(e.g. ~\cite{LBH96}; \cite{BMS00}; \cite{VRH01}).  Its binding energy onto
a dust grain covered with water ice is thought to be a factor of about
1.5 smaller than that of CO ($E_{\rm D}$(O) $\sim$ 800 K; Tielens \&
Allamandola 1987).  We have therefore run several models which include atomic
oxygen in the chemistry, with different values of $E_{\rm D}$(O), and 
determine the $\chi^2$ and column density
profiles in the same fashion as before.    These results are shown
in Fig. ~\ref{fbfito} and correspond to models 3 and 4 of
Table ~\ref{tbfit}.  One sees that model 3 in particular gives
a reasonable fit although the \DCOP \ column density profile is
still somewhat deviant.  
Both models 3 and 4 require an oxygen 
binding energy of $\sim$ 600 K.

Using the ``best fit'' models in Tab.~\ref{tbfit}, namely Models 1 and 3,
we determined the variation of the electron fraction with cloud
radius $r$.  Fig.~\ref{ffrac_abun} shows the fractional abundance of
electrons, \HCOP , \DCOP , \NTHP , and \NTDP , as well as the depletion
factor $f_{\rm D}$, as a function of $r$.  
Model 3 (the best fit model) predicts a 
[\NTDP ]/[\NTHP ] aboundance ratio equal to $\sim$ 0.4 at radius 
$r$ $\sim$ 2500 AU.  
Both models show
$x(e)$ between $\sim$ 10$^{-9}$ at $r$ $\sim$ 2500 AU ($n(\rm H_2)$
$\sim$ 10$^6$ \percc ) and $\sim$ 10$^{-8}$ at $r$ $\sim$ 10$^4$ AU
($n(\rm H_2)$ = a few $\times$ 10$^4$ \percc ).  In these models,
the dependence of $x(e)$ upon  gas density $n(\rm H_2)$ is approximately
given by :
\begin{eqnarray*}
x(e) & = & 6.3 \times 10^{-6} \times n(\rm H_2)^{-0.64},
\end{eqnarray*}
in Model 1, and
\begin{eqnarray*}
x(e) & = & 5.2 \times 10^{-6} \times n(\rm H_2)^{-0.56},
\end{eqnarray*}
in Model 3.  These estimates
are roughly an order of magnitude lower than the
standard relation between $x(e)$ and $n(\rm H_2)$
($x(e)$  =  1.3$\times 10^{-5} \times
n(\rm H_2)^{-0.5}$ ;  \cite{M89}).  In other words, our results are
compatible with electron fractions only a factor of order  3 higher
than the lower limits based on the inferred \HCOP \ abundance
in the nucleus of L1544.
Such low electron fractions are linked to 
the depletion of metals which, being positively charged,
 directly affect the electron budget of cloud cores. We use an 
initial abundance of metals 10 times lower than in the standard relation 
quoted above and this causes the difference in the coefficient (see 
Sect.~\ref{sdescription}).  The
difference in the exponent is to be attributed to the further metal 
depletion inside the core.  If larger initial metal abundances are
used, $x(e)$ increases at the outer edges of the
core, but it does not significantly change at the core center,
where metal abundances are negligible because of depletion. In this case,
a steeper slope in the $x(e)-n(\MOLH )$ relation is obtained.
We caution  that these estimates have large uncertainty as the
differences in the models of Tab.~\ref{tbfit} shows. Our approach
to inferring molecular
ion densities as a function of radius is questionable and the real
geometry is non spherically symmetric. There are additionally
many uncertainties in the chemistry. Nevertheless, we conclude
tentatively that the ``standard relationship'' may give a considerable
over--estimate of the electron fraction in L1544.  This has consequences
for the ambipolar diffusion timescale.

In the central portion of L1544, ionization degrees
of the order of 10$^{-9}$ imply an ambipolar diffusion time
scale of roughly 2.5$\times$10$^{13}$ $x(e)$ = 2.5$\times$10$^4$ yrs
(see \cite{S78} and \cite{SAL87}), comparable to the free--fall
time scale at densities of $\sim$ 10$^6$ \percc . This is consistent
with the view that within the central region, (see ~\cite{CB01},
\cite{CB00}), conditions are ``super--critical''
 and the core is rapidly developing towards a
situation where it will collapse. However, the ``caveats'' noted
above mean that one needs to test this conclusion with observations
capable of better delineating the structure of the high density
nucleus of L1544.

\subsubsection{Chemical problems with the best fit model}

There are some problems related with the existence of atomic
oxygen in the gas phase which need to be clarified. In our models, 
the large abundance of
gaseous O is due to the low value of $E_{\rm D}$(O) and to the
cosmic ray desorption mechanism which allows a prompt return of
this species to the gas phase, once adsorbed onto grain surfaces.
We are neglecting here the fact that the reactions of atomic oxygen
with \HTHP \ necessary to reduce deuterium fractionation will also
produce water and OH in the gas phase which can then deplete out
onto grains, although only a small fraction of O ($<$ 0.01) will be 
lost due to this process (e.g. ~\cite{LBH96}).   An important 
consequence of the \HTHP + O reaction is that \HTHOP \ becomes the most 
abundant molecular ion in the depleted core ($x({\rm H_3O^+})$ $\sim$ 
10$^{-9}$ at $n({\rm H_2})$ $\sim$ 10$^{6}$ \percc 
as also found by ~\cite{AOI01}).  
However, the destruction rate of atomic
oxygen through reactions with \HTHP \ ($r_{\rm H_3^+}$) 
is significantly slower than the O accretion rate onto 
dust grains ($r_{\rm dust}$) \footnote{At a density of $\sim$ 10$^6$ --
10$^5$ \percc , our best fit model predicts $x(\HTHP )$ $\sim$ 1 -- 2
$\times$ 10$^{-10}$, so that
$r_{\rm H_3^+}$ = $k_{\rm H_3^+} \times x(\HTHP ) \times n({\rm H_2})$
$\sim$ 1 -- 2 $\times$ 10$^{-14}$ \percc s$^{-1}$, where the rate coefficient 
$k_{\rm H_3^+}$ $\sim$ 10$^{-9}$ cm$^3$ s$^{-1}$. For the O accretion rate,
assuming a unity sticking coefficient and dust grains of 10$^{-5}$ cm in 
size,  $r_{\rm dust}$ $\sim$ 10$^{-17}$ $n({\rm H_2})$ \percc , implying
$r_{\rm dust}$/$r_{\rm H_3^+}$ $\sim$ 50 -- 100 at $n({\rm H_2})$ = 10$^5$ --
10$^6$ \percc , respectively.}.  
This means that if the accreted O will not be completely processed 
onto grain surfaces, through, say, hydrogenation leading to water, 
between successive cosmic ray bombardments, we expect a fraction of 
free oxygen to be maintained on the gas phase.  Proving the validity 
of these assumptions is outside the scope of this paper but we consider 
briefly the likely chemistry of atomic O on grain surfaces.

Following \cite{HH93}
(see ~\cite{LJO85}), if the major source of nonthermal
grain heating is due to  Fe nuclei with energies 20--70 MeV
nucleon$^{-1}$, the time interval between successive
cosmic ray impacts is
about $10^6$ years.  If the fractional abundance of atomic
oxygen is $\sim$ 10$^{-4}$ and $n(\rm H)$ $\sim$ 1 \percc , as gas phase
models of dense clouds predict at steady state
for $\zeta = 10^{-17}$ s$^{-1}$ (e.g. ~\cite{LBH96}),
$T$ = 10 K, and $n(\rm H_2)$ $\sim$ 10$^5$ \percc , the O and H accretion
rates are $\sim$ 4$\times$10$^{-5}$ s$^{-1}$ and 1$\times$10$^{-5}$ s$^{-1}$,
respectively, i.e. between two cosmic ray bombardments
about 10$^9$ O and 3$\times$10$^8$ H atoms can accrete on the surface of a
grain.  Assuming that hydrogen can quickly move on the grain surface
(from the laboratory work of ~\cite{KFB99}, the H diffusion rate
on silicate grains is about 2$\times$10$^{-4}$ s$^{-1}$, much larger than
the c.r. heating rate),  a fraction between 15\% and 30\%  of surface O
will be hydrogenated and transformed in OH and H$_2$O.
The rest of the O atoms may
form O$_2$, but observations have shown that this is probably not the main
repository of oxygen either in the solid (\cite{VEB99}) or in the gas phase
(\cite{GMB00}).  It is thus possible that a large fraction of atomic
oxygen remains unreacted on the grain before the next cosmic ray will 
release it back in the gas phase.  

It is relevant in this contest that the detailed gas--grain
chemical--dynamical model of L1544, presented by \cite{AOI01} (which includes 
depletion onto dust grains but no surface processing of accreted species) 
predicts large O fractions across the core (from $\sim$ 10$^{-4}$ at 
$n({\rm H_2})$ $\sim$ 10$^{5}$ \percc \ to a few $\times$ 10$^{-5}$ at
10$^6$ \percc , similar to our results). This is in their ``fast collapse'' model, 
which fits the observed CCS and CO distributions.   Our 
model is crude compared with that of \cite{AOI01}, but nevertheless 
both models reproduce the observed features quite well, proving the validity 
of our chemical assumptions in Sect.~\ref{sdescription}.  
There are some differences between the two 
models concerning the prediction of \NTHP \ and \NTDP \ abundances. In \cite{AOI01}, 
the calculated column densities of the two species are smaller than those
observed by about an order of magnitude, perhaps due to uncertainties
in the \MOLN \ formation mechanism.  We ``avoid'' this problem by simply starting
the chemistry with all the available nitrogen already in the form of 
gas phase \MOLN \ (Sect.~\ref{sdescription}),  thus reproducing
the observed \NTHP \ and \NTDP \ column density variations (see Model 3 in 
Fig.~\ref{fbfit}).

\section{Conclusions}
\label{sdiscussion}

 An important result of this study is
  that \NTHP \ shows more deuterium fractionation
  towards the dust emission peak of L1544 than
 \HCOP (models without depletion gradients expect
 [\DCOP]/[\HCOP] = [\NTDP]/[\NTHP] ~\cite{RC01}). We suspect
 that the high degree of deuteration observed in ammonia
 (see ~\cite{TRF00},~\cite{RTC00}) in some cores can be best
 explained by a model similar to that which
 we have adopted. Ammonia like \NTHP \ is easily
 produced from \MOLN \ (see ~\cite{RC01}) and is therefore likely
 to be relatively abundant in the depleted region. The compact
 nature of ``ammonia cores'' is naturally explained if this
 is the case (\cite{TMC01}).

 We have attempted to obtain density estimates for the regions within
 the L1544 core giving rise to the observed \DCOP \ and \NTHP (\NTDP )
 emission. We obtain in this way values between $10^5$ and $10^6$
 \percc \ consistent with the general idea that we are observing
 selectively the high density layers. The results from \DCOP \
 suggest densities as high as $10^6$ \percc \ consistent with the
 values inferred from dust emission and absorption.  However,
 lower values are obtained based upon our \NTHP \ and \NTDP \ data.
 This is surprising and we are presently unsure of the explanation.
 The observed spatial distribution suggests that \NTHP \ and
 \NTDP \ reside in higher density layers than \DCOP .
 The reasons for this discrepancy  could have to do
 with the isothermal assumption which we have made based upon our
 \CEIO \ and \AMM \ results.
Attempts to simulate the
 dust emission from L1544 (~\cite{ZWG01},~\cite{ERSM01} ) have shown
 that there is probably a subtantial fall off in dust temperature
 from the center to the edge with values of order 7~K in the nucleus.
 At densities above $10^5$ \percc (\cite{KW84}) gas
 and dust temperatures are probably coupled and hence there may
 be a gradient in the gas temperature too (see also \cite{G01}).
 This should be taken into account in future studies.

  We have developed a crude model of the ion--chemistry in the
  core of L1544. This simulates the observed depletion and can
  reproduce the observed dependence of
  the column densities of species such as \NTHP ,\NTDP ,
  \HCOP , and \DCOP \ as a function of offset.  The main
  advantage of this is that it allows us in objective fashion to
  consider the relative contributions to observed column
  densities from the high density depleted nucleus and
  the lower density foregrond (background) gas. There are large
  uncertainties in both the input to the chemistry and the
  process of inferring radial dependences of molecular abundances
  from the observations.  One interesting result is that we get
  the best fit to the observed deuterium fractionation in models in
  which atomic oxygen is allowed to remain with an abundance of order
  $10^{-4}$ in the gas phase.  There are some problems associated
  with this but one needs more detailed models of both the surface
  and gas phase chemistry to test this. It does have an observational
  consequence which is that \HTHOP \ becomes a major ion towards the
  dust peak, in agreement with numerical chemical models of \cite{AOI01}.
  
  Applying this model, we find that the ionization degree in
  L1544 is likely to be an order of magnitude smaller than estimated
  using ``canonical formulae'' existing in the literature, mainly 
because of the reduced metal abundances. The result
  is a tentative finding because we have difficulty in fitting the
  observed abundances as a function of offset in L1544 and
  in particular the observed degree of D fractionation. However,
  our estimates for ionization degree do not vary greatly when one
  compares the different models which give an adequate fit to the
  observed column density distributions. We conclude
   that in the case of L1544, the ionization degree is
  lower and hence the ambipolar 
 diffusion timescale is shorter.  Our estimated timescale is of the same 
order as the free-fall time consistent with the idea that the nucleus of
the L1544 core is undergoing dynamical collapse.  

A final point to emphasize is that in
models without atomic oxygen, or where O also experiences strong depletion, 
the main ion in the highly depleted region is 
  \HTHP , so that the mean mass of positive ions decreases from $\sim$ 30
(in the ``canonical'' undepleted case, where the representative ion is 
\HCOP ) to 
$\sim$ 3.  This has the consequence of further reducing the ambipolar time
scale, due to a drop in the drag coefficient,  by a factor of 7--8 
(D. Galli, priv. comm.).  

  Other cores may yield different results and the   effects of the
  possible temperature gradient mentioned above need to be taken into
  account.  Finally, it will be important in future
  studies to take into account a more realistic geometry
  (non spherically symmetric) than we have done.

\acknowledgements
The authors are grateful to the referee, Neal Evans, for useful comments
and suggestions. PC and CMW
 wish to acknowledge travel support from ASI Grants 66-96,
 98-116, as well as from the MURST project ``Dust and Molecules
in Astrophysical Environments''.  We are grateful to Daniele Galli
for providing the subroutine for the calculation of the electron
fraction.

\clearpage

\appendix

\section{Column density determination}
\label{acolumn}

In addition to the LVG calculations, we also used the following expressions
to determine the total column density.  For optically thick transitions:
\begin{eqnarray}
N_{\rm TOT} & = & \frac{8 \pi^{3/2} \Delta v}{2 \sqrt{ln 2}
 \lambda^3 A} \frac{g_l}{g_u} \frac{\tau}{1 - exp(-h \nu/k T_{\rm ex})}
 \frac{Q_{\rm ROT}}{g_l exp(-E_{l}/k T_{\rm ex})},
\label{anthick} 
\end{eqnarray}
where $\Delta v$ is the line width, $\lambda$ and $\nu$ are the wavelength 
and frequency of the observed transition, respectively, $A$ is the Einstein
coefficient, $g_l$ and $g_u$ are 
the statistical weight of the lower and upper levels, $\tau$ is the optical 
depth, $h$ is Planck constant, $T_{\rm ex}$ is the excitation temperature 
(assumed the same for all rotational levels), $Q_{\rm ROT}$ is the partition
function, $E_l$ is the energy of the lower level, and $k$ is Boltzmann 
constant.  For linear molecules (as those observed in this paper):
\begin{eqnarray}
Q_{\rm ROT} & = & \sum_{J=0}^{\infty} (2J+1) exp[-E_J/(k T)] \\
E_J & = & J (J+1) h B, 
\end{eqnarray}
where $J$ is the rotational quantum number, and $B$ is the 
rotational constant.   
For rotational transitions with hyperfine structure (e.g. \NTHP (1--0) and 
\NTDP (2--1)), $\tau$ refers to the 
total optical depth (given by the sum of the peak optical depths of all the 
hyperfine components), and $\Delta v$ to the intrinsic line width. 
The error on $N_{\rm TOT}$ is given by propagating the errors on $\Delta v$,
$\tau$, and $T_{\rm ex}$ in eqn.~\ref{anthick}. 

If a line is optically thin and all the rotational levels are 
characterized by the same excitation temperature $T_{\rm ex}$,
the expression of the total column density becomes:

\begin{eqnarray}
N_{\rm TOT} & = & \frac{8 \pi W}{\lambda ^3 A} \frac{g_l}{g_u} 
 \frac{1}{J_{\nu}(T_{\rm ex}) - J_{\nu}(T_{\rm bg})} 
 \frac{1}{1 - exp[-h\nu /(k T_{\rm ex})]} 
 \frac{Q_{\rm ROT}}{g_l exp[-E_l/(k T_{\rm ex})]},
\label{anthin}
\end{eqnarray}   
where $W$ is the integrated intensity of the line ($W$ = 
$\sqrt{\pi}/(2 \sqrt{ln 2})$ $\times$ $\Delta v T_{\rm mb}$, for a 
gaussian line, with $T_{\rm mb}$ $\equiv$ main beam brightness temperature),
$J_{\nu}(T_{\rm ex})$ and $J_{\nu}(T_{\rm bg})$ are the 
equivalent Rayleigh--Jeans excitation and background temperatures.
The integrated intensity of each line has been measured by integrating
over a velocity range determined in the following way: (i) sum all the 
spectra in the map; (ii) find the $rms$, or the 1 $\sigma$ level of the
noise in the off--line channels of the sum spectrum; (iii) include in the 
velocity range for integration (at zero level) all the channels in the 
line which are above the 1 $\sigma$ level determined in point (ii).  For the 
calculation of the total column density only those positions with 
$W/\sigma_W$ $\geq$ 3 have been considered ($\sigma_W$ is the error on 
$W$, determined by the expression $\sigma_W$ = $\Delta v_{\rm res}$ $\times$
$rms$ $\times$ $\sqrt{N_{\rm ch}}$, where $\Delta v_{\rm res}$ is the spectral
resolution in \kms, and $N_{\rm ch}$ is the number of channels in the 
integrated area).  
The error on $N_{\rm TOT}$ is simply given by $\sigma_{N_{\rm TOT}}$ = 
$\sigma_W$ $\times$ $N_{\rm TOT}/W$, 

The parameters used to determine $N_{\rm TOT}$ for each species are 
listed in Table~\ref{amol}.

\begin{table}
\centering
\caption{Molecular parameters used to calculate $N_{\rm TOT}$} 
\vspace*{2mm}
\begin{tabular}{ccccc}
\hline
Transition & $B$ & $\mu$ & $\Delta v_{\rm res}$ & $N_{\rm ch}$ \\
         &  MHz & Debye & \kms & \\
\hline
\NTDP (2--1) &   38554.719 &  3.4 &  0.038 &     275 \\
\HCEIOP (1--0) & 42581.21  &  3.9 &  0.034 &      17 \\
\HTHCOP (1--0) & 43377.32  &  3.9 &  0.034 &      25 \\
\HCSEOP (1--0) & 43528.933 &  3.9 &  0.067 &       6 \\
\NTHP (1--0) &   46586.867 &  3.4 &  0.063 &      75 \\     
\CEIO (1--0) &   54891.420 &  0.11 & 0.027 &      33 \\
\CSEO (1--0) &   56179.990 &  0.11 & 0.026 &      72  \\
\DTHCOP (2--1) & 35366.712 &  3.9  & 0.041 &      14  \\
\DCOP (2--1) &   36019.76  &  3.9  & 0.041 &      22  \\
\DCOP (3--2) &   36019.76  &  3.9  & 0.054 &      13  \\
\hline
\end{tabular}
\label{amol}
\end{table}

\subsection{\DCOP \, and \HCOP }
\label{adcop}

As we saw in Sect.~\ref{sdensity}, an LVG statistical equilibrium code
has been used to estimate number density and \DCOP \ column density
from observed \DCOP (3--2) and (2--1) lines. 
To quantify the uncertainties in column density estimates from 
statistical equilibrium (SE) calculations, due to the unknown 
\DCOP \ fractional abundance and H$_2$ number density, we first 
determined \DCOP \ column densities from the SE code and a least square
method applied to model calculated data points. To do this, the SE
program was run for different values of H$_2$ number density (between 10$^3$ 
and 10$^6$ cm$^{-3}$) and \DCOP  \, fractional abundances (between 10$^{-12}$ 
and 10$^{-9}$).  Each run furnishes a value of $T_{\rm ex}$, $\tau$, $T_{\rm 
mb}$ of the first four rotational transitions of \DCOP, together with 
the \DCOP \ column density divided by line width.  Secondly, a least--square 
method was applied to these model data points to find general expressions for 
estimating $N(\DCOP )$ once the brightness temperature 
$T_{\rm mb (2-1)}$ of the J = 2-1 line and the 
$W(\DCOP (3-2))/W(\DCOP (2-1))$ ($\equiv$ $R_{\rm W}$) integrated intensity 
ratio are known from observations.  However, this gives $N(\DCOP )$ values 
with associated errors greater than 30\% ($N/\sigma_{\rm N}$ $<$ 3), so we
decided to first determine the excitation temperature ($T_{\rm ex (2-1)}$)
and the optical depth ($\tau_{(2-1)}$) of the J = 2--1 line using the 
least--square method, which gives:
\begin{eqnarray}
T_{\rm ex (2-1)} & = & a_1 + a_2 log R_{\rm W}  + 
 a_3 T_{\rm mb}(2-1) \label{eqtex}\\
log \, \tau_{(2-1)} & = & b_1 + b_2 (log R_{\rm W})^2 
 + b_3 log (T_{\rm mb}(2-1)) \label{eqtau}, 
\end{eqnarray}
with $a_1$ = 11.3\p0.1, $a_2$=10.3\p0.2, $a_3$=-0.05\p0.01, $b_1$=-0.2\p0.1,
$b_2$=2.0\p0.5, $b_3$=1.0\p0.1.  Secondly, we assumed $T_{\rm ex}$ = 
$T_{\rm ex (2-1)}$ for all rotational levels, and finally calculate $N$ 
using expressions (\ref{anthin}) or (\ref{anthick}) in the appendix, 
depending on the value of the optical depth ($\tau_{(2-1)}$ $<$ or $>$ 0.5, 
respectively).  Only positions inside the half maximum 
contour of the \DCOP (2--1) integrated intensity map have been included
in this calculation, to exclude low sensitivity spectra from the analysis. 
The particular form of these expressions 
was chosen because of the relatively low $\chi^2$. 
If from (\ref{eqtau}) $\tau$/$\sigma_{\rm \tau}$ $<$ 3, 
$\tau_{(2-1)}$ was estimated from the radiative transfer equation 
\begin{eqnarray}
\tau & = & -log \left[ 1 - \frac{T_{\rm mb}}{J_{\nu}(T_{\rm ex}) - 
 J_{\nu}(T_{\rm bg})} \right] \label{eqtau2},
\end{eqnarray}
where $J_{\nu}(T_{\rm ex})$ and $J_{\nu}(T_{\rm bg})$ are the equivalent 
Rayleigh-Jeans excitation and background temperatures, and $T_{\rm ex}$ is
given by eqn.~(\ref{eqtex}).  Finally, if the 
resultant $N/\sigma_{\rm N}$ $<$ 3, 
$T_{\rm ex}$ was fixed to 7 K, its mean value\footnote{$T_{\rm ex}$ 
drops from $\sim$ 9 K towards the map peak, at offset (20, -20), 
and two adjacent positions to $\la$ 7 K 
at larger distances from the peak.}, and $\tau$ estimated 
from (\ref{eqtau2}).  $T_{\rm ex}$ = 7 K was also assumed for all the 
positions outside the half maximum contour of the \DCOP (2--1) integrated 
intensity map, where eqns.~(\ref{eqtex}) and (\ref{eqtau}) cannot be 
applied because of the poor $R_{\rm W}$ estimate. 
 In the particular case of the (20, -20) offset position, 
the optically thin \DTHCOP (2--1) line
was used to estimate the \DCOP \ column density, assuming the same excitation
temperature found for the \DCOP (2--1) line.   {\it 
We note that the different
approaches used in estimating column densities do not change the results by 
more than a factor of 2}.  However, it is important to estimate the 
optical depth of the line before determining the column density.  
For example, in the case of offset (20, -20) 
where the \DCOP (2--1) line has one
of the largest values of optical depth ($\sim$ 2), 
(i) the SE 
calculation together with equation (\ref{anthick}) give $N(\DCOP )$ = 
(4\p1)$\times$10$^{12}$ \cmsq , (ii) the use of $T_{\rm ex}$ = 7 K and 
 eqn.(\ref{eqtau2}), gives $\tau$ = 1.7\p0.2 and $N(\DCOP )$ = 
(2.0\p0.2)$\times$10$^{12}$ \cmsq, and (iii) the use of \DTHCOP (2--1) leads 
to $N(\DCOP )$ = (4.0\p0.3)$\times$10$^{12}$ \cmsq.  If the 
optical depth of the \DCOP (2--1) line is not taken into account, one obtains 
$N(\DCOP )$ = (9.6\p0.2)--(10.2\p0.2)$\times$10$^{11}$ \cmsq, 
if $T_{\rm ex}$ = 7 - 9 K, respectively.  
  
For the determination of the \HCOP \ column density, the excitation temperature
was assumed equal to that found for \DCOP.  We used the optically thin 
\HCEIOP (1--0) line, if available, and the \HTHCOP (1--0) line in all 
other positions.  The passage from $N(\HCEIOP )$ and $N(\HTHCOP)$ to
$N(\HCOP )$ was made by multiplying the former quantities by 560 and 
77, respectively, based on the [$^{16}$O]/[$^{18}$O] and 
[$^{12}$C]/[$^{13}$C] abundance ratios in the local ISM (\cite{WR94}).

The optical depth of \HTHCOP (1--0) was calculated from 
eqn.~(\ref{eqtau2}).  As in the case of 
\DCOP, eqn.(\ref{anthin}) or (\ref{anthick}) was used, depending on the value 
of $\tau$. It is 
interesting to note that the resultant $\tau$ in many positions is less than 
0.5, i.e. \HTHCOP (1--0) is mostly optically thin and thinner than the 
\DCOP (2--1) line in the central part of the core. However, 
the profile of \HTHCOP (1--0) suggests the existence of 
absorption by a low density foreground layer (see ~\cite{TMM98},
\cite{WMW99}) which may cause a decrease in the brightness temperature
and, consequently, in the line optical depth (see eqn.~\ref{eqtau2}).     
The three hyperfine components of 
the \HCSEOP (1--0) transition detected at offset (20, -20), and shown in 
\cite{DCC01} together with their recent measurement
in the laboratory, allowed us to check the 
thickness of the \HCEIOP (1--0) line.  Assuming $T_{\rm ex}$ = 9 K,
the excitation temperature found from \DCOP \ data at the same position,  
we find $N(\HCEIOP )/N(\HCSEOP )$ = 4.2\p0.4, quite close to the 
[$^{18}$O]/[$^{17}$O] ratio (= 3.65; \cite{P81}) expected in optically thin 
conditions.  

The determination of the \HCOP \ column density starting from \HTHCOP \
is affected by an addictional complication, i.e. the presence of 
$^{13}$C fractionation in cold and dense gas (e.g. ~\cite{SA84}) 
due to the exothermic reaction:
\begin{eqnarray*}
^{13}{\rm C}^+ + ^{12}{\rm CO} & \rightarrow & ^{12}{\rm C}^+ + ^{13}{\rm CO} 
	+ \Delta E,
\end{eqnarray*}
with $\Delta E/k$ = 35 K (Watson 1977).  In optically thin conditions,
the $N(\HTHCOP )/N(\HCEIOP )$ column density ratio should be equal to 
the product of [$^{13}$C]/[$^{12}$C] and [$^{16}$O]/[$^{18}$O] abundance 
ratios ($\sim$ 7).  In L1544, $N(\HTHCOP )/N(\HCEIOP )$ 
ranges from 3 (at offset [80, -80]) to 7 (at offset [40, -80]) with 
an average value of 4\p1.  It is thus 
always lower than the local interstellar medium value, suggesting that
$^{13}$C fractionation is not significantly affecting our conclusions. 
However, given that \HTHCOP (1--0) is probably affected by foreground 
absorption, 
it is extremely hard to estimate the effects of $^{13}$C fractionation. 
From the current data we determined the $W[\HTHCOP (1-0)]/W[\HCEIOP (1-0)]$
integrated intensity ratio and found an average value of 4\p1,  
smaller than the value expected in optically thin conditions and 
no $^{13}$C fractionation.  This result suggests that optical depth effects on the
\HTHCOP (1--0) line are in any case predominant.

\subsection{CO}

For the CO column density, we used 
\CSEO (1--0), when available, and \CEIO (1--0) in the other 
positions assuming $T_{\rm ex}$ = 10 K (Sect.~\ref{stemperature}), 
[$^{17}$O]/[$^{16}$O] =
2044, and [$^{17}$O]/[$^{16}$O] = 560 (\cite{WR94}, \cite{P81}).  
\CSEO (1--0) 
has (well resolved) hyperfine components with relative intensity ratios 
consistent with optically thin emission, whereas \CEIO (1--0) has a moderate 
optical depth ($\tau$ between 0.5 and 1), based on 
the comparison between the $W(\CEIO (1-0))/W(\CSEO (1-0))$ integrated 
intensity ratio and the [$^{18}$O]/[$^{17}$O] 
abundance ratio in the local ISM.  Following \cite{MLB83}, we derive:
\begin{eqnarray}
\frac{T_{\rm mb}[\CEIO (1-0)]}{T_{\rm mb}[\CSEO (1-0)]} & = & 3.65 \times 
 	\left( \frac{ 1 - exp(-\tau_{18})}{\tau_{18}} \right), \label{eqtau18}
\end{eqnarray}
where $T_{\rm mb}(i)$ is the main beam brightness temperature of line
$i$ and $\tau_{18}$ is the \CEIO (1-0) optical depth.  Once $\tau_{18}$ 
is known from the above equation, the excitation temperature of the 
\CEIO ~line can be determined from the radiative transfer equation:
\begin{eqnarray}
T_{\rm ex} & = & \frac{h \nu/k}{ln \left( \frac{h \nu}{k J_{\nu}(T_{\rm ex})} 
 + 1 \right) },
\end{eqnarray}
where $\nu$ is the frequency of the \CEIO (1--0) line.  
  The variation in $T_{\rm ex}$ from point to 
point in the map is not significant, considering the errors, and we assumed a 
constant value of $T_{\rm ex}$ (= 10 K) given by the weighted mean of the 
excitation temperature in the observed positions. 
  As in the case of $N(\DCOP )$ and 
$N(\HTHCOP )$, when $\tau$ $>$ 0.5, the \CEIO ~column density has been 
calculated using eqn.~(\ref{anthick}). 

\subsection{\NTHP \, and \NTDP }
\label{an2hp}

The observed \NTHP (1--0)  and \NTDP (2--1) lines present hyperfine structure 
due to the interaction between the molecular electric field gradient and
the electric quadrupole moments of the two nitrogen nuclei.  The 
$J$ = 1$\rightarrow$0 line of \NTHP \ is splitted in seven components
(see \cite{CMT95}), whereas \NTDP (2--1) has 38 hyperfines which
partially overlap because of line broadening.  

We first determined the intrinsic
linewidth, total optical depth, and excitation temperature from the 
hfs fitting procedure 
applied to high sensitivity spectra (selected by requiring $W$/$\sigma_W$ 
$>$ 20, $W$ being the intensity integrated under the seven hyperfine 
components).  In fact, the optical depth determination becomes more and more
uncertain with increasing spectral rms, which can alter the relative 
intensities of the components and thus crucially affect $\tau$ 
estimates.  Therefore it is important to have clear detections 
of the seven components before attempting an hfs fit.  

However, even in the presence of high sensitivity spectra one can encounter 
problems if the transition is very thick ($\tau$ $\geq$ 30) because the CLASS 
fit procedure is limited to $\tau$ values smaller than 30.  In these cases, it 
is convenient to estimate the \NTHP (1--0) optical depth from the thinnest 
component (the lowest frequency $J_{\rm F_1,F}$ = 1$_{1,0}$ $\rightarrow$ 
0$_{1,1}$ line, see ~\cite{CMT95}) assuming a certain 
value of $T_{\rm ex}$ (e.g. the mean value, $<T_{\rm ex}>$, which in 
L1544 is 5 K), and using eqn.~\ref{eqtau2}.  The hfs fit gives $\tau$ 
= 30 - with a suspiciously low associated error ($\sim$ 0.1) - in four 
positions of the L1544 \NTHP (1--0) map (offsets [40, -40], [20, -40], 
[40, -60], and [20, -60]).  For these spectra, the use of eqn.~\ref{eqtau2} 
with $T_{\rm ex}$ = 5 K confirms the presence of high optical depth 
($\leq$ 30).  In these cases, we used the integrated 
intensity of the ``weak'' and moderately thick ($\tau_{1,0-1,1} \leq 1$) 
$F_1,F$ = 1,0 $\rightarrow$ 1,1 component to determine the total \NTHP \ 
column density from eqn.(\ref{anthin}).  Although this may underestimate
the total column density by a factor of 
$\sim$ $\tau/(1 - exp(-\tau{1,0-1,1}))$ $\leq$ 1.6, the uncertainty associated
on the above estimate of the total optical depth is also large (about 20\%),
and does not take into account of the possible presence
of excitation anomalies, as found by ~\cite{CMT95}.    
For consistency, the use of the $F_1,F$ = 1,0 $\rightarrow$ 1,1 component to 
estimate the total \NTHP \ column density has been extended to all other 
positions where this hyperfine is well detected.  

For those \NTHP (1--0) spectra which cannot be hfs fitted because of 
low sensitivity, and where the ``weak'' component is not visible, 
we assumed optically thin conditions, $T_{\rm ex}$ 
= 5 K, and determined $N(\NTHP )$ using eqn.~\ref{anthin}.  

As we already pointed out at the beginning of this section, 
the $J$ = 2 $\rightarrow$ 1 transition of \NTDP ~is splitted in 38 hyperfine 
components and only in one position (the integrated intensity map peak, where
the \NTDP (2--1) observations have been repeated several times to check 
the system) it has been possible to determine the total optical depth and 
excitation temperature from the hfs fit ($\tau_{\rm tot}$ = 4.9\p0.6, 
$T_{\rm ex}$ = 4.9\p0.8 K).   In all the 
other positions we (i) assumed optically thin conditions, (ii) estimated the 
integrated intensity below the hyperfines ($W$) with its uncertainty
($\sigma_{\rm W}$),  (iii) excluded all spectra with $W$/$\sigma_W$ $<$ 3, 
(iv) fixed $T_{\rm ex}$ at 5 K,  and  (v) use eqn.~(\ref{anthin}) to estimate
the total column density.

\clearpage

\begin{center}
\begin{deluxetable}{cccccc}
\tablewidth{0pc}
\footnotesize
\tablecaption{Parameters of the ``best fit'' models}
\tablehead{
 \colhead{Parameters} & \colhead{Standard} & \colhead{Model 1} &
 \colhead{Model 2} & \colhead{Model 3} & \colhead{Model 4} }
\startdata
\hline
Density Dist.     &   WMA   &  WMA & BAP  &  WMA  &  BAP  \nl
$\zeta$(s$^{-1}$) &  1.3$\times$10$^{-17}$ & 5.0$\times$10$^{-19}$ &
 1.3$\times$10$^{-18}$ & 6.0$\times$10$^{-18}$ & 5.0$\times$10$^{-18}$ \nl
$E_{\rm D}$(CO)[K] & 1210 & 960 & 1210 & 1210 & 1210 \nl
$E_{\rm D}$(N$_2$)[K] & 787 & 600 & 600 & 787 & 787 \nl
$E_{\rm D}$(O)[K] & \nodata & \nodata & \nodata & 600 & 600 \nl
$a_{\rm min}$(cm) & 1$\times$10$^{-6}$ & 2.5$\times$10$^{-5}$ &
 1.0$\times$10$^{-5}$ & 5.0$\times$10$^{-6}$ & 1.0$\times$10$^{-6}$ \nl
$S$ & 1.0 & 0.1 & 0.1 & 1.0 & 0.1 \nl
$\chi^2$ & \nodata & 4.7 & 6.7 & 1.6 & 5.4 \nl
\enddata
\label{tbfit}
\end{deluxetable}
\end{center}

\clearpage


\figcaption[density.ps]{Logarithmic values of the H$_2$ number density,
$n(\rm H_2)$, superposed upon the half maximum contours of the \DCOP (2--1)
(thin curve), the \DCOP (3--2) (dashed curve) integrated intensity maps,
and the 1.3mm continuum dust emission (dotted curve)
convolved to a 22 \arcsec ~beam (circle in the bottom right).
The grey scale shows the \DCOP (3--2) integrated intensity map (levels 50,
70, and 90\% of the peak, 0.84 K \kms at offset [20, -20]).  The density
values in the figures are based on statistical equilibrium calculations
(see text). \label{fnh2}}

\figcaption[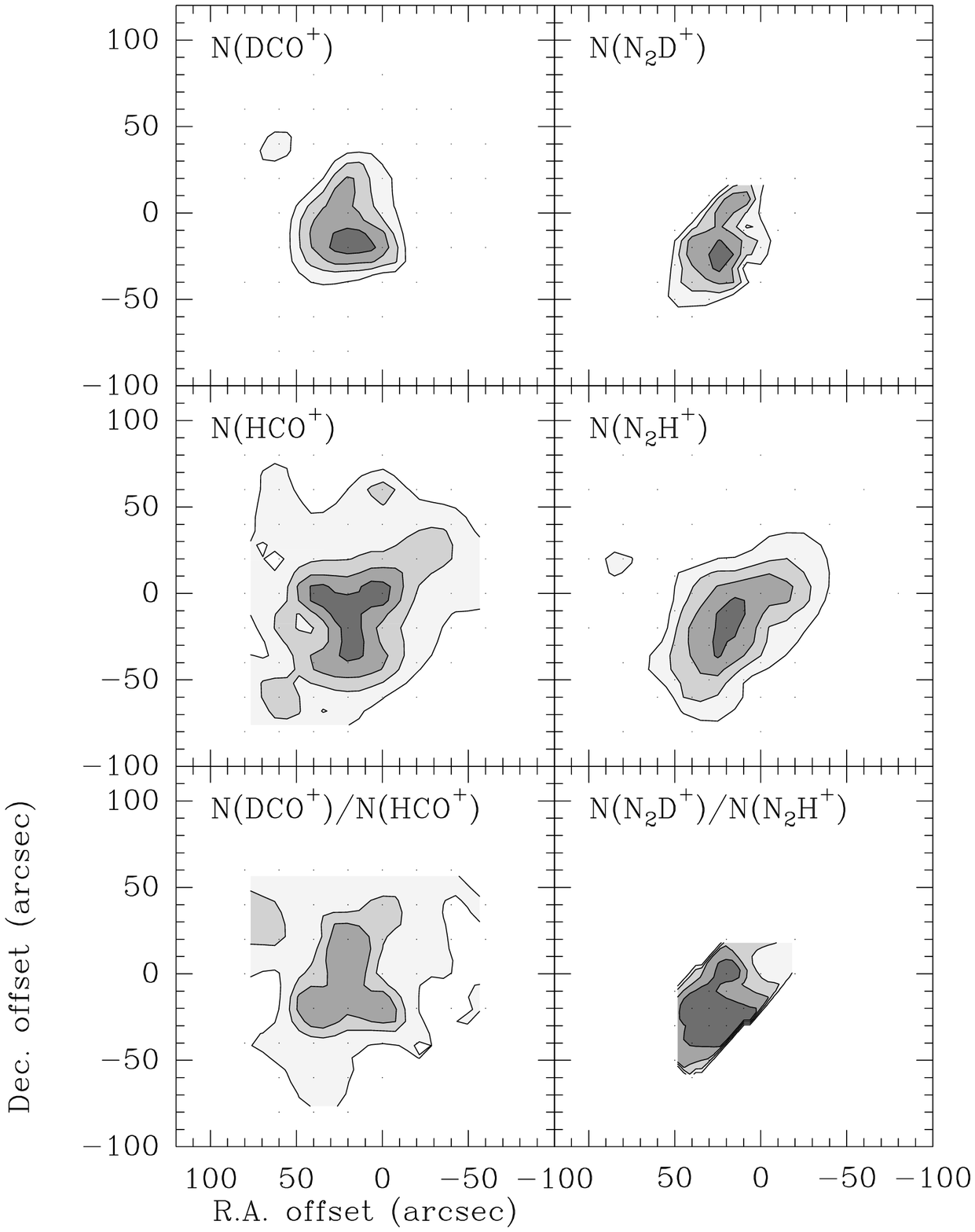]{Map of the depletion factor $f_{\rm D}$ (grey scale)
superposed to the 1.3mm continuum dust emission map of WMA,
smoothed at a resolution of 22\arcsec \ (dotted contours).
$f_{\rm D}$ ranges between 1 (at the core edges) and 10.8 (at offset [24,
-34]); the grey scale (from light to dark) shows parts in the core with
$f_{\rm D}$ = 1, 3, 5, 7, and 9. Dotted contours levels represent 30, 50,
70, and 90\% of the 1.3mm map peak (225 mJy/22\arcsec at offset [26, -21]).
The distance between the two peaks is about half a beam. \label{fdep}}

\figcaption[n_dco+_hco+.ps]{({\it top left}) \DCOP \ column density map;
contour levels are between 30 and 90\%, in steps of 20\% of the map peak
(= 4.0$\times$10$^{12}$ \cmsq , at [20, -20]).
({\it center left}) \HCOP \ column density map; contour levels are as in the
previous panel and the
map peak is at offset (0, 0) (1.1$\times$10$^{14}$ \cmsq). ({\it bottom left})
$N(\DCOP )/N(\HCOP )$ column density ratio map.
Contour levels are 10, 30, and 50\% of the map peak (0.06\p0.02 at offsets
[20, 20] and [40, -20]). The difference between the four positions
inside the 50\% contour level is not significant, taking into account
the associated errors.  ({\it top right}) \NTDP \ column density map
(contours range between 30 and 90\%, in steps of 20\%, of the map peak 
(= 4.4$\times$10$^{12}$ \cmsq , at [20, -20]). ({\it center right}) 
\NTHP \ column density map; contour levels are as in the previous panel 
and the map peak is at offset (20, -10) (2.0$\times$10$^{13}$ \cmsq ).
({\it bottom right}) $N(\NTDP )/N(\NTHP )$ column density ratio map;
contour levels are 10, 30, 50, 70\% of the peak (0.26 at [40, -20]).  
\label{dco-hco}}

\figcaption[fig_standard.ps]{Dependence on radius of the depletion
factor for CO ($f_{\rm D}$) and N$_2$ (top panels), electron fraction and
various ionic species (bottom panels) for the ``standard model'' (see
Tab.~\ref{tbfit} for input parameters) using the density distribution found by
WMA (left panels) and BAP (right panels).
\label{fstandard}}

\figcaption[figmol_dist.ps]{Column density (filled circles) versus impact 
parameter for
(from top to bottom): CO, \HCOP , \DCOP , \NTHP , and \NTDP .   White
symbols are average column densities inside the corresponding
bin, with the only exception of the value at $b$ = 0 (see text).
\label{fmol_dist}}

\figcaption[figwt_bac.ps]{Column density profiles predicted by the
``best fit'' models (solid curves) and deduced from observations
(dotted curves).  Left panels refer to models with a volume density profile
from WMA (Model 1), whereas models shown in right
panels assume the L1544 density profile from BAP (Model 2).
\label{fbfit}}

\figcaption[figwt_bac_o.ps]{Column density profiles predicted by the
``best fit'' models with atomic oxygen (solid curves) and observed
(dotted curves).  As in Fig.~\ref{fbfit}, left panels refer to models
with volume density profile from  WMA (Model 3), whereas
right panels show models with density profiles from BAP (Model 4).
The inclusion of atomic oxygen in the gas phase furnishes a better
agreement with observed molecular ion column densities at the dust 
peak and with observed column density profiles. 
\label{fbfito}}

\figcaption[frac_abun.ps]{Radial profiles of fractional abundances
of electrons, \HCOP , \DCOP , \NTHP , and \NTDP \, predicted by
Model 1 (top) and 3 (bottom; see Tab.~\ref{tbfit}).  Dashed curves
represent the depletion factor. Electron fractions are about one 
order of magnitude lower than deduced from ``standard'' chemical 
models without depletion.  This affects the dynamical evolution 
of prestellar cores by shortening the ambipolar diffusion time 
scale. 
\label{ffrac_abun}}

\plotone{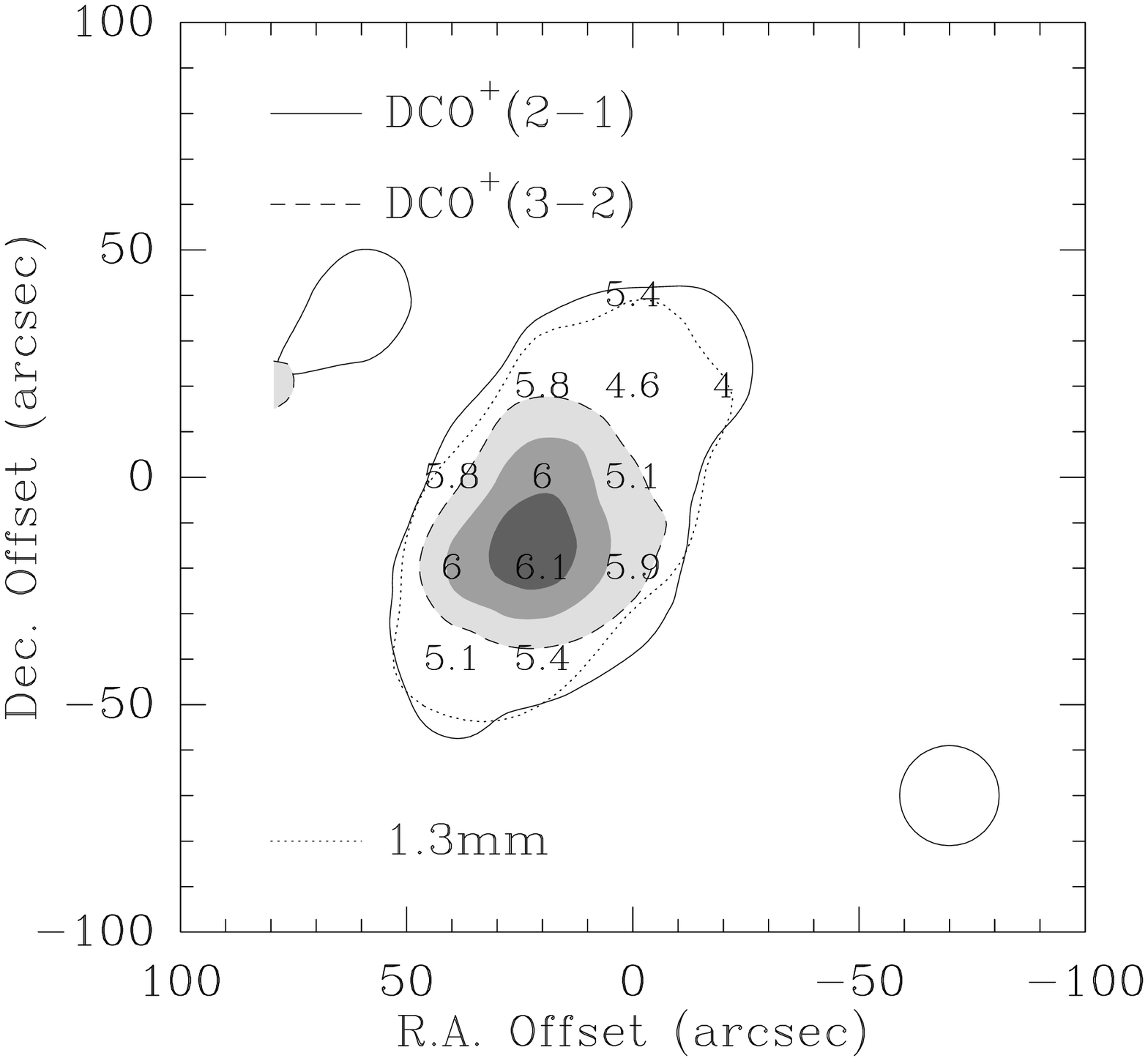}
Fig.1

\plotone{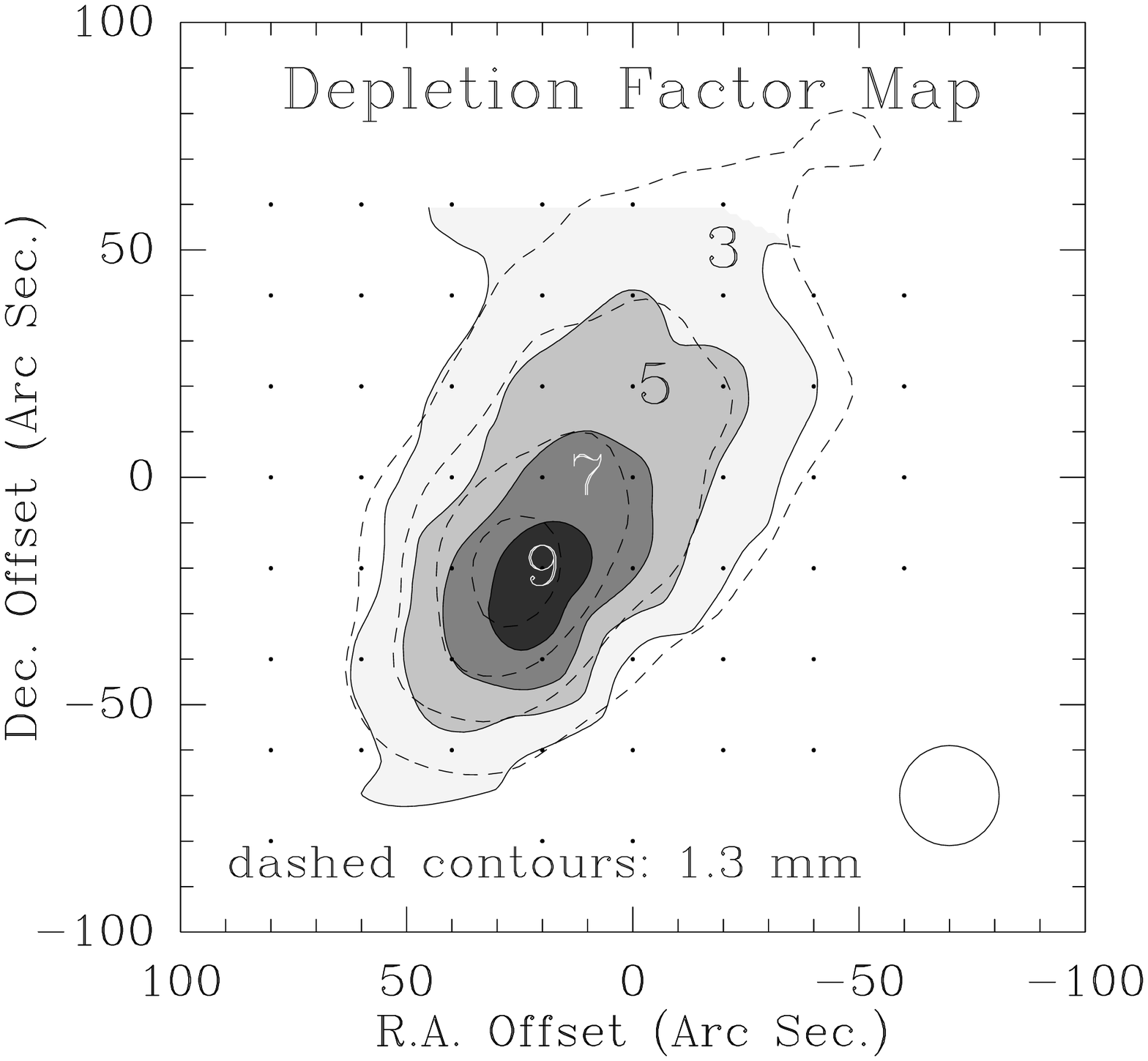}
Fig.2

\plotone{fig3.ps}
Fig.3

\plotone{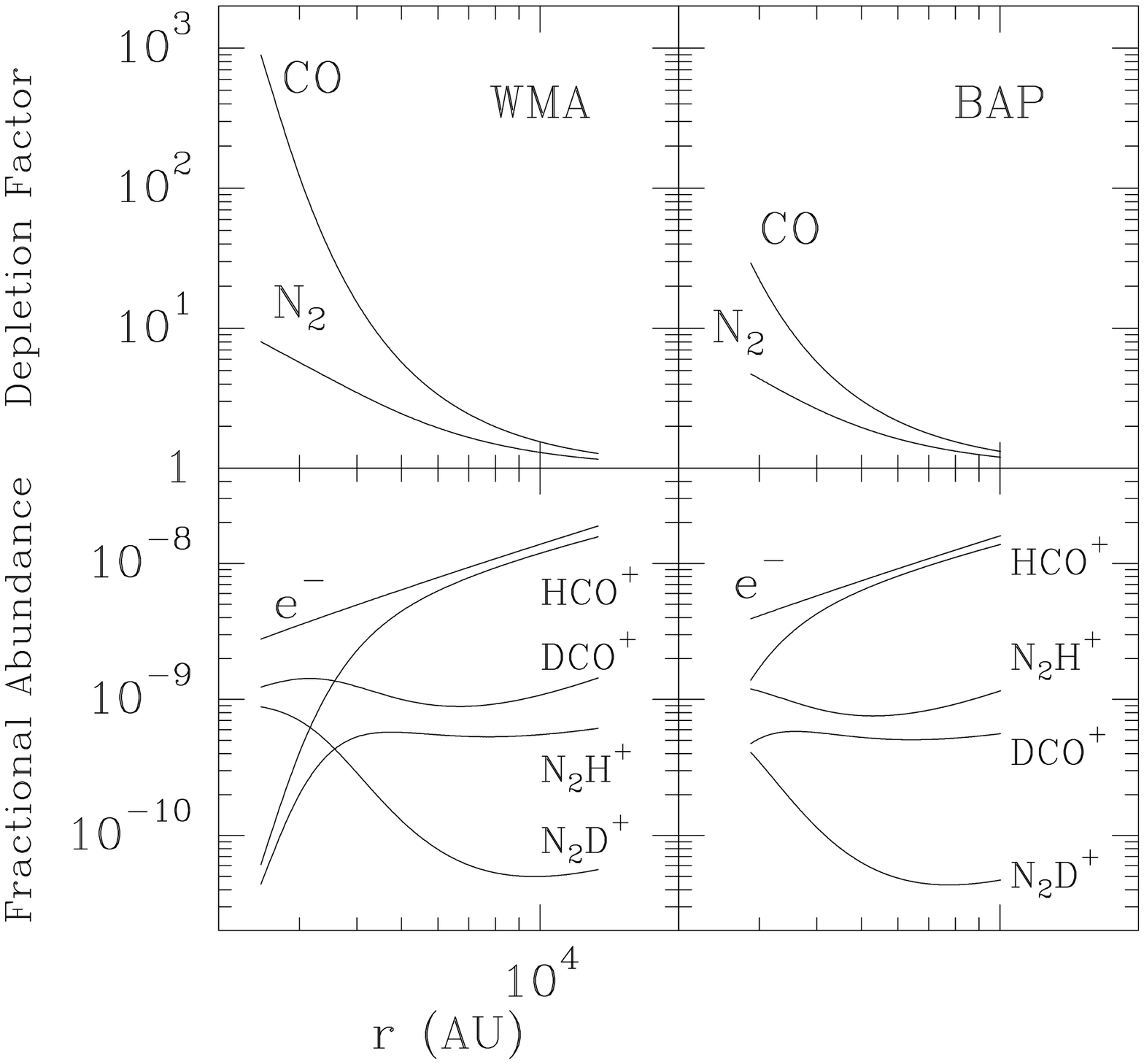}
Fig.4

\plotone{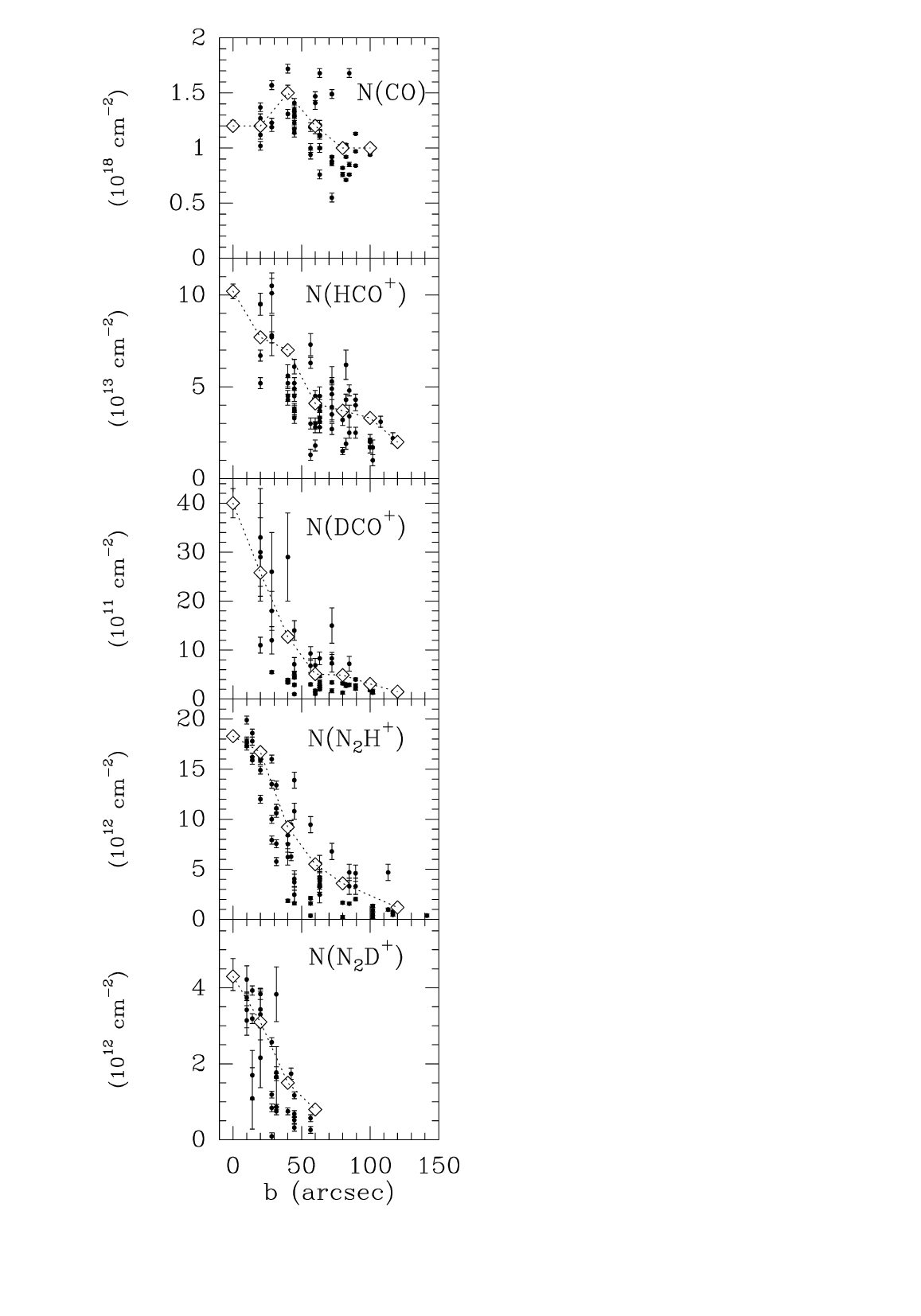}
Fig.5

\plotone{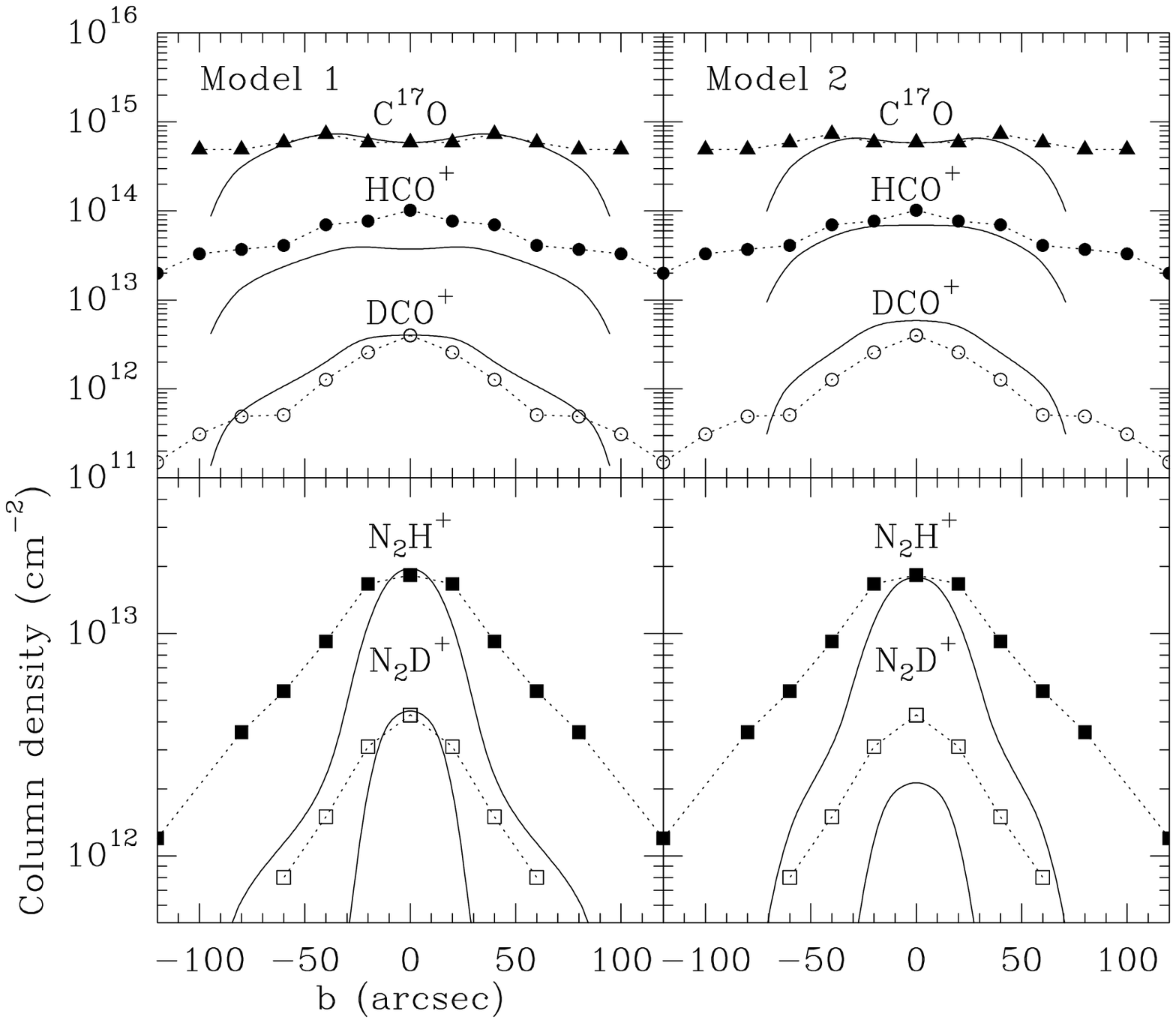}
Fig.6

\plotone{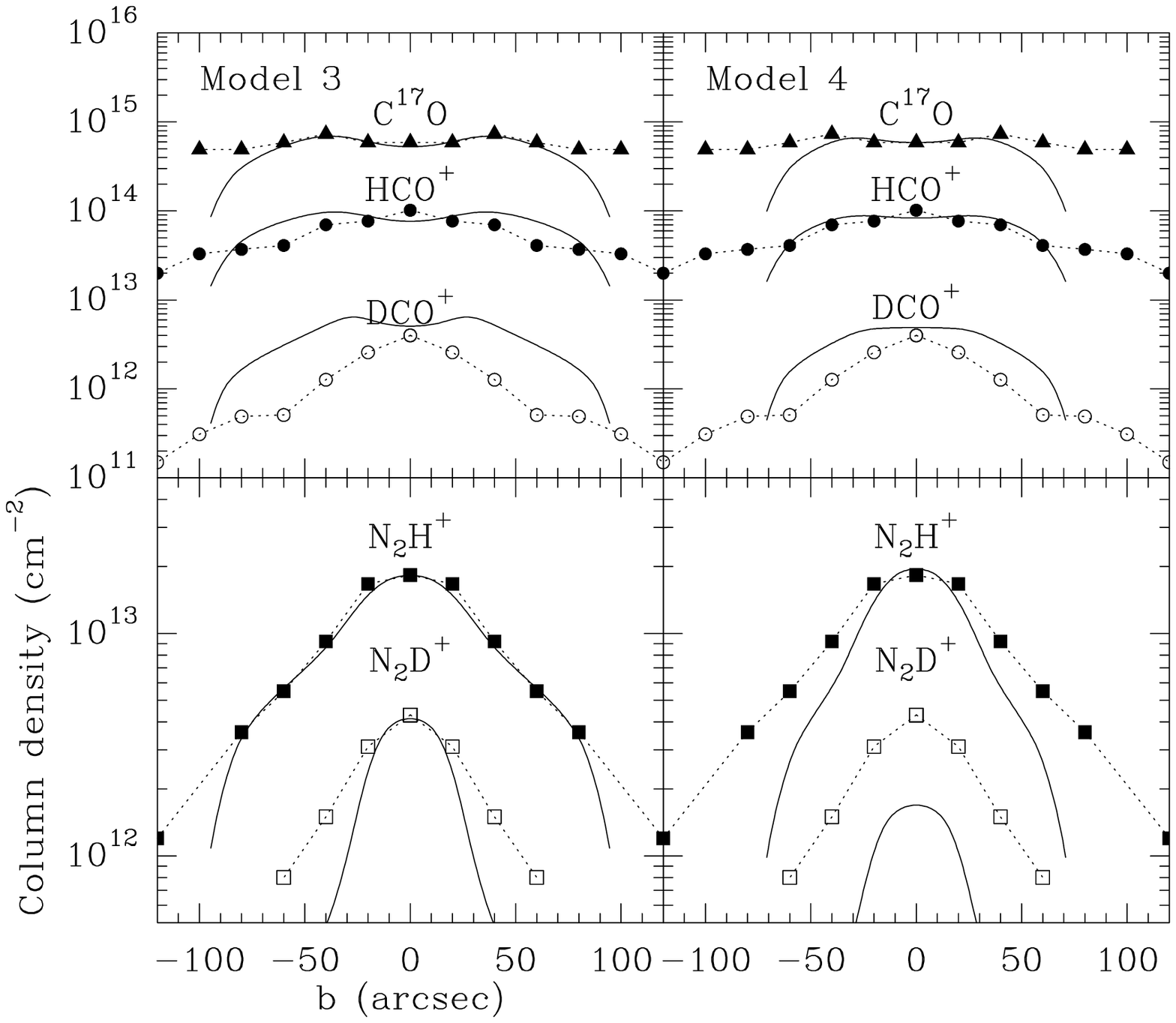}
Fig.7

\plotone{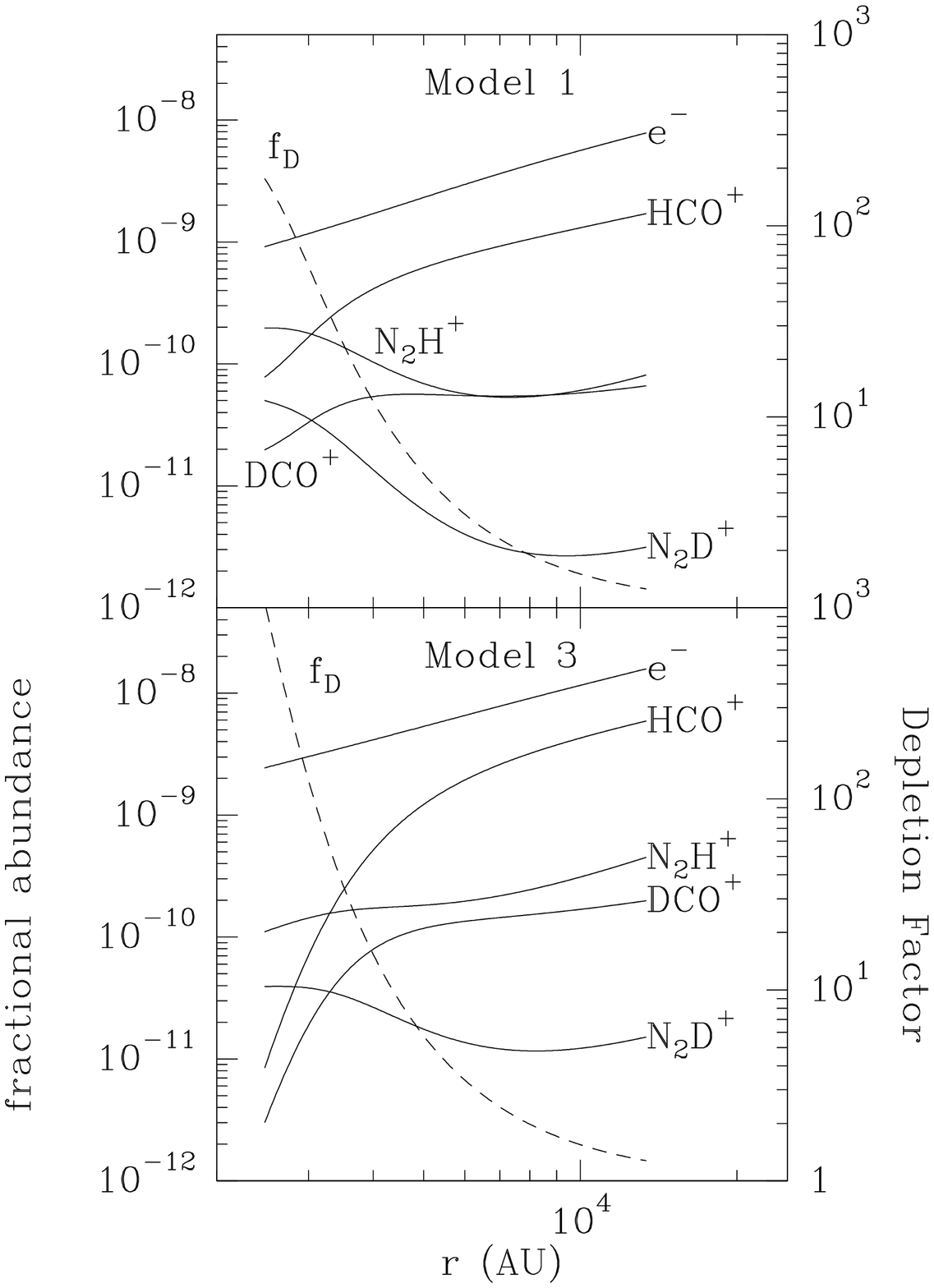}
Fig.8



\end{document}